\documentclass[a4paper,UKenglish]{lipics-v2016}

\usepackage{amssymb}
\usepackage{amsfonts}
\usepackage{mathtools}
\usepackage{url}
\usepackage{graphicx}

\usepackage{thmtools} 
\usepackage{thm-restate}

\usepackage{color}
\usepackage[dvipsnames]{xcolor}
\usepackage{graphics,hyperref}
\usepackage[numbers]{natbib}
\usepackage{tikz}
\usepackage{tkz-tab}
\usetikzlibrary{arrows,positioning,automata}
\usepackage{tkz-graph}
\usepackage{enumerate}
\usepackage[subrefformat=parens,labelformat=parens]{subcaption}
\usepackage{array}

\usepackage{authblk}

\newtheorem{fact}{Fact}

\newcommand{\MCadj}{\mathcal{M}_{adj}}
\newcommand{\MCany}{\mathcal{M}_{any}}
\newcommand{\MCanyrev}{\mathcal{M}_{any}^{*}}

\usetikzlibrary{shapes,snakes}
\hypersetup{colorlinks=true,linkcolor=blue,citecolor=blue,filecolor=blue,urlcolor=blue}
\GraphInit[vstyle=Dijkstra]

\newcommand{\states}{{
		\Vertex[x=0,y=1.5]{abc}
		\Vertex[x=6.5, y=1.5]{cba}
		\Vertex[x=2, y=3]{bac}
		\Vertex[x=2, y=-0]{acb}
		\Vertex[x=4.5, y=3]{bca}
		\Vertex[x=4.5, y=0]{cab}
	}}

	\newcommand{\abcbac}{{\Edge[label=$a-b$](abc)(bac);}}
	\newcommand{\bacabc}{{\Edge[label=$b-a$](bac)(abc);}}
	\newcommand{\abcacb}{{\Edge[label=$b-c$](abc)(acb);}}
	\newcommand{\acbabc}{{\Edge[label=$c-b$](acb)(abc);}}
	\newcommand{\bacbca}{{\Edge[label=$a-c$](bac)(bca);}}
	\newcommand{\bcabac}{{\Edge[label=$c-a$](bca)(bac);}}
	\newcommand{\bcacba}{{\Edge[label=$b-c$](bca)(cba);}}
	\newcommand{\cbabca}{{\Edge[label=$c-b$](cba)(bca);}}
	\newcommand{\cabcba}{{\Edge[label=$a-b$](cab)(cba);}}
	\newcommand{\cbacab}{{\Edge[label=$b-a$](cba)(cab);}}
	\newcommand{\cabacb}{{\Edge[label=$c-a$](cab)(acb);}}
	\newcommand{\acbcab}{{\Edge[label=$a-c$](acb)(cab);}}
	\newcommand{\acbbca}{{\Edge[label=$a-b$, style={bend left = 40}
			](acb)(bca);}}
	\newcommand{\bcaacb}{{\Edge[label=$b-a$, style={bend left = 40}
			](bca)(acb);}}
	\newcommand{\abccba}{{\Edge[label=$a-c$](abc)(cba);}}
	\newcommand{\cbaabc}{{\Edge[label=$c-a$](cba)(abc);}}
	\newcommand{\baccab}{{\Edge[label=$b-c$, style={bend left = 40}
			](bac)(cab);}}
	\newcommand{\cabbac}{{\Edge[label=$c-b$, style={bend left = 40}
			](cab)(bac);}}

	\tikzset{EdgeStyle/.style={->,thick,
			double          = blue!50,
			double distance = 1.5pt}}
	
	\newcommand{\edgeMany}{\tikzset{EdgeStyle/.style={->,thick,
				double          = red!50,
				double distance = 1.5pt}}}
	\newcommand{\edgeManyrev}{\tikzset{EdgeStyle/.style={->,thick,
				double          = green!50,
				double distance = 1.5pt}}}	

\newtheorem{observation}{Observation}
\usepackage{aliascnt}
\usepackage{cleveref}
\newaliascnt{lemma}{theorem}
\aliascntresetthe{lemma}
\crefname{lemma}{lemma}{lemmata}
\Crefname{lemma}{Lemma}{Lemmata}

\title{\bf Sort well with energy-constrained comparisons}
\author{Barbara Geissmann}
\author{Paolo Penna}
\affil{Department of Computer Science, ETH Zurich, Zurich, Switzerland}

\begin{document}
\maketitle
\begin{abstract}
	We study very simple sorting algorithms based on a \emph{probabilistic} comparator model. In our model, errors in comparing two elements are due to (1) the energy or effort put in the comparison and (2) the difference between the compared elements. Such  algorithms keep comparing pairs of randomly chosen elements, and they  correspond to  Markovian processes. The study of these Markov chains reveals an interesting phenomenon. Namely, in several cases, the algorithm which repeatedly compares  only \emph{adjacent} elements is \emph{better} than the one making arbitrary comparisons: on the long-run, the former algorithm produces sequences that are ``better sorted''.  The analysis of the underlying Markov chain poses new interesting questions as the latter algorithm yields a non-reversible chain and therefore its stationary distribution seems difficult to calculate explicitly.  
\end{abstract}
\section{Introduction}
Suppose one has to sort a number of elements by making pairwise comparisons, but sometimes the result of a comparison is \emph{uncorrect}. Sometimes the errors are unavoidable, and sometimes they are deliberately introduced in order to save other important resources. For example, with \emph{probabilistic CMOS} it is possible to \emph{trade energy for errors}, that is,  one can reduce the energy spent for a single operation but this will increase the probability of incorrect response (see the survey of \citet{PalemA13}). Errors also occur in measurements that require high precision (where a small noise can affect the result), or judgment made by individuals (who naturally tend to make small mistakes). One can envision the following situation:
\begin{enumerate}
	\item It is ``easier'' to compare two elements if they differ ``a lot'', while errors are more likely when they are ``very close''. \label{comp:close}
	\item If we are able or willing to spend more energy (effort) on a single comparison, then we can increase the probability of getting the correct result. \label{comp:energy}
\end{enumerate}
Given this scenario, one would like to design a strategy (algorithm) which sorts the elements nearly correctly. The following is thus a natural question:
\begin{quote}
	\emph{Which strategies (algorithms) perform better?}
\end{quote}
This question has been already addressed under various models of errors (see e.g. \cite{QuicksortPhaseTrans,braverman2008noisy,recursiveMergeSort}). The purpose of this work is to study this question under a new model which captures the two features above (\Cref{comp:close,comp:energy}).

\subsection{Our contribution}
In this work we propose to look at extremely simple \emph{sorting} algorithms on what we call a \emph{probabilistic comparator} model. These algorithms can be studied through the lens of \emph{Markov chains} whose analysis reveals interesting properties regarding their behavior. %

\paragraph*{Probabilistic comparator model}
We introduce a simple model in which the probability that a comparison between two elements (numbers) is correct depends on two factors: (1) how different the two elements are and (2) the ``effort'' or ``energy'' spent to make the comparison. Intuitively speaking, the more energy the more accurate is the comparison, meaning that even very similar elements (small difference) can be distinguished.  More precisely, for an energy parameter $\lambda \geq 1$, a comparison between two numbers $a$ and $b$ is correct with probability 
\[
p_{ab}:=\frac{\lambda^{b-a}}{\lambda^{a-b}+\lambda^{b-a}}, 
\] 
where $b$ is the biggest between $a$ and $b$. One should think about a single comparison as a measurement of two quantities, which sometimes can be erroneous especially when the difference is small. All comparisons (including those involving the same two numbers) are independent and performed with the same parameter $\lambda$. That is, we consider \emph{independent} errors in which the error probabilities are described by the formula above.

\begin{remark}
	Our model is inspired by the classical models in \emph{statistical physics} \cite{mezard2009information} and in \emph{game theory} \cite{blumeGEB93} where $\lambda=e^{1/noise}$. In the latter applications, $1/noise$ corresponds to the ``rationality level'' of players, i.e., their ability to distinguish between strategies with similar payoffs. In statistical physics, the $noise$ parameter is the ``temperature'' of the system, and high temperature corresponds to highly disordered configurations.
	To some extent, this model can be seen as an abstraction of the \emph{probabilistic CMOS} technology which allows to trade energy for correctness \cite{PalemA13}. As our  model attempts to abstract from hardware-specific construction details, it necessarily introduces certain simplifying assumptions. Arguably, the major of these assumptions is that the probability of errors depends uniquely on the difference between the two numbers to be compared, and not on their actual values. On the other hand, the  parameter $\lambda$ captures the property that, by increasing the energy per single operation, the probability of correct comparisons increases in some way (depending on the hardware).
\end{remark}

\begin{figure}
	\centering
	\includegraphics[width=.5\textwidth]{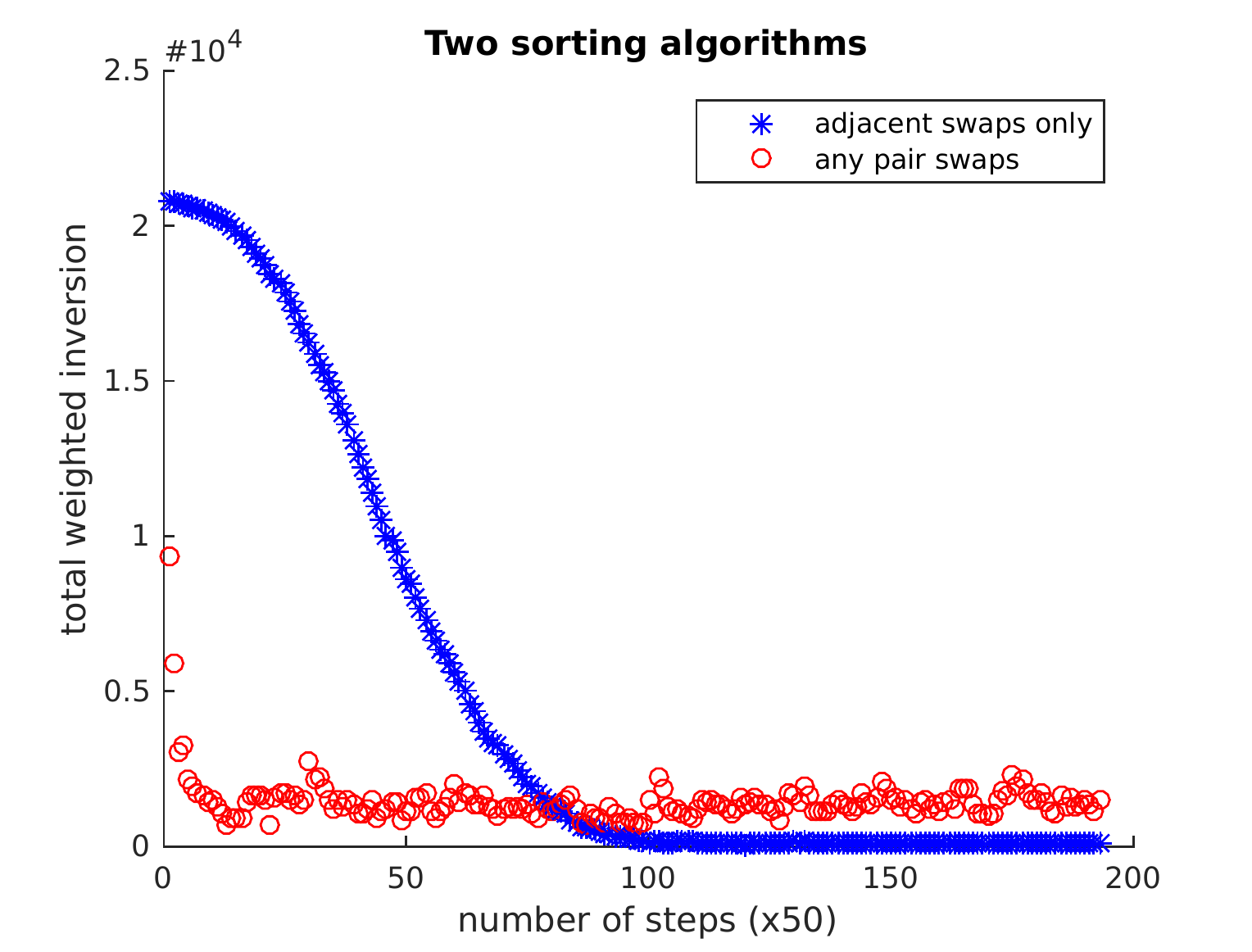}
	\caption{Comparison of two simple algorithms for sorting: on the long run the algorithm doing only adjacent comparisons gives a better result compared to the one doing all comparisons. The input sequence is $\{50,49,\dots,1\}$ while the sorted sequence is $\{1,2,\ldots 50\}$. In this experiment we set $\lambda = e^{\frac{1}{5}}$, though the same behavior occurs for essentially any fixed $\lambda$ (see Section~\ref{sec:experiments} for the experiments).} 
	\label{fig:experiment}
\end{figure}

\paragraph*{Sort by random comparisons}
When sorting several elements, one performs several comparisons for a certain number of steps and then returns the resulting sequence. 
Consider the following two simple algorithms:
\begin{itemize}
	\item \emph{Any pairs swaps}. Compare two randomly chosen elements of the sequence.
	\item\emph{Adjacent swaps}. Compare a randomly chosen element with the next one in the sequence.
\end{itemize}
Surprisingly enough, Figure~\ref{fig:experiment} shows that the algorithm with only adjacent swaps gives \emph{better} results  after a certain number of comparisons are made (intuitively, the $y$-axis measures the ``disorder'' in the current sequence in each algorithm).  Note that the initial input sequence is irrelevant as long as we consider sufficiently many steps of such algorithms, a property that can be formally captured by viewing these algorithms as Markov chains. 

\paragraph*{Algorithms and Markov chains}
The two algorithms above (and others) can be viewed as Markov chains whose stationary distribution describes the output if we let them run long enough. We believe that these processes are interesting by themselves (for instance, they can be seen as variants of other well-studied processes -- see Section~\ref{sec:related}), and they pose new questions on how to analyze them. Specifically, the adjacent swaps algorithm corresponds to a \emph{reversible} Markov chain $\MCadj$ and its stationary distribution has a simple closed formula of the form
\begin{equation}
\label{eq:prob_sequence}
\pi(s) \propto \lambda^{-2w(s)}
\end{equation}
where $w()$ is what we call a \emph{total weighted inversion} of sequence $s$, 
\[w(s) := \sum_{i<j \colon s_i>s_j}{s_i-s_j}\enspace ,\]
a measure of its ``distance'' from the correctly sorted sequence. Intuitively, inversions involving very different elements count more than inversions of almost identical elements.  In contrast, the algorithm with arbitrary (random) pair comparisons corresponds to a \emph{non-reversible} chain $\MCany$, and therefore the analysis of its stationary distribution is considerably more complicated. We also provide a variant of $\MCany$ in which  comparisons of non-adjacent pairs are done \emph{multiple times}: For example, if numbers $a$ and $b$ are two positions away in the current sequence (say $a=s_i$ and $b=s_{i+2}$) then we perform \emph{two} comparisons and accept to swap them only if \emph{both} of them tell to do so. This third chain $\MCanyrev$ has the same stationary distribution of $\MCadj$. Therefore, one can see this ``careful swapping'' rule as a way to fix the algorithm doing naively arbitrary swaps. The analysis of this chain $\MCanyrev$ is based on the \emph{Kolmogorov reversibility criterion}.

 Figure~\ref{fig:experiment}, and all experiments we made on various input sequences, suggest that  $\MCadj$ yields \emph{better sorted} sequences  than $\MCany$, though the latter chain \emph{converges faster} to its stationary distribution.
 We study both the \emph{mixing time} and the properties of the \emph{stationary distribution}, like the probability of returning the sorted sequence. 
\begin{itemize}
	\item In Section~\ref{sec:binary}, we consider the case of \emph{binary sequences}, where each element in the sequence is either $a$ or $b$. We show that the mixing time of $\MCadj$ is $O(n^2)$, while for $\MCany$ it is  $O(n\log n)$, or even linear if the number of occurrences of $b$ is constant, for every $\lambda>1$.
	\item In Section~\ref{sec:stationary-compare} we study the probability that the chains return the sorted sequence at stationary distribution. We show that  $\MCadj$ is better than $\MCany$ when sorting \emph{three arbitrary elements}. This result is based on the  \emph{Markov chain tree theorem} and it is the most involved in this section. Similar results hold also for sorting arbitrary long sequences with a \emph{single outlier}, that is, binary sequences with a single element $b>a$ and many $a$'s. Here the analysis shows a quantitative difference between the two chains (cf. Theorem~\ref{th:outlier}). Note that, all these results apply also to $\MCanyrev$ in place of as $\MCadj$, since they have the same stationary distribution.  
\end{itemize}

\subsection{Related work}\label{sec:related}
Stochastic models of the form \eqref{eq:prob_sequence} are very common in statistics and, in particular, \citet{Mal57} was among the firsts to consider such models in the context of permutations: There the weight function $w()$ is a  suitable distance function which comes from probabilities $p_{ab}$ of ranking $a$ before $b$. In that sense, our model  is a special case of Mallows' model, though the procedure of \cite{Mal57} is different: One makes all pairwise comparisons at once until a consistent result is obtained. Our probabilistic comparator is also a special case of \citet{BraTer52} where the probability $p_{ab}$ of ranking $a$ before $b$ is of the form $\frac{w_a}{w_a+w_b}$. 

Several restrictions on  $p_{ab}$ have been studied for the natural Markov chain which makes only adjacent comparisons. The classical card shuffling problem corresponds to the \emph{unbiased} version of this chain in which all probabilities $p_{ab}$ equal $1/2$, for which \citet{Wil04} proved that this chain is rapidly mixing and gave a very tight bound. A similar problem is the uniform sampling of partial order extensions, which corresponds to probabilities $p_{ab}$ being $1/2$ or $1$ and $p_{ba}=1-p_{ab}$. For the latter, \citet{BubDye99} showed that this chain is also rapidly mixing. \citet{Benetal05} proved rapidly mixing for the \emph{constant biased} case, that is, when every comparison is correct with some fixed probability $p>1/2$, independently of the compared elements: $p_{ab}=p>1/2$ for all $a<b$. The mixing time of \emph{biased} comparisons has been studied by \citet{Bhaetal13} under two comparison models called ``choose your weapon'' and ``league hierarchies'': In the first model $p_{ab}$ depends only on the largest between $a$ and $b$, while in the second model all numbers are the leaves of some tree and $p_{ab}$ depends only on the least common ancestor of $a$ and $b$.  Note that our model does not fall in either class even for the case of only three distinct elements.

\citet{DiaRam00}  studied a different type of chains called systematic scan algorithms: for the \emph{unbiased} case, they proved that  $n$ of such scan operations are sufficient to reach the stationary distribution.
 
\subsection{Preliminary definitions on Markov chains}
In this section we introduce some of the definitions on Markov chains used throughout this work (for more details see \citet{LevPerWil09}). A Markov chain over a finite state space $S$ is specified by a transition matrix $P$, where $P(s,s')$ is the probability of 
moving from state $s$ to state $s'$ in one step. The $t^{th}$ power of the transition matrix gives the probability of moving from one state to another state in $t$ steps. All chains studied in this work are ergodic meaning that they have a unique \emph{stationary distribution} $\pi$: for any two states $s$ and $s'$,  $\lim_{t\rightarrow \infty} P^t(s,s')=\pi(s')$.  

We will use the definition of a \emph{reversible} Markov chain, also called \emph{detailed balanced condition}: If the transition matrix $P$ admits a vector $\pi$ such that $\pi(s) P(s,s') = \pi(s')P(s',s)$ for all $s$ and $s'$, then $\pi$ is the \emph{stationary distribution} of the chain with transitions $P$. 

\renewcommand{\P}{\mathbf{P}}

An equivalent characterization of reversible chains is given by looking at cycles over the states. For any subset $\Gamma \subseteq S\times S$ of transitions (pairs of states of the chain), define the associated probability as the product of all these transitions in the chain, \begin{equation}
	\label{eq:prob_transition_set}
	\P(\Gamma):= \prod_{(x,y)\in \Gamma} P(x,y) \enspace .
\end{equation}
Let $\Gamma^{-1}$ denote the reversed edges in $\Gamma$, that is, $\Gamma^{-1}:=\{(y,x)|\ (x,y) \in \Gamma\}$. The \textit{Kolmogorov reversibility criterion} \cite[][]{Kel11} says that a chain is reversible if and only if for any cycle $C$, 
\begin{equation}
\label{eq:Kolmogorov_criterion}
\P(C)=\P(C^{-1}) \enspace .
\end{equation}
For the sake of clarity, we sometimes denote cycles as $C=s^{1}\rightarrow s^{2}\rightarrow \cdots \rightarrow s^{\ell} \rightarrow s^{1}$ and the corresponding reversal by  $C^{-1}= s^{1}\leftarrow s^{2}\leftarrow \cdots \leftarrow s^{\ell} \leftarrow s^{1}$. 

The stationary distribution of any (even non-reversible) Markov chain can be computed by looking at the probabilities of all directed trees rooted at some state. More formally, let $\mathcal T(s)$ be the set of all directed trees rooted at state $s$, that is, from every other state there is a path towards $s$ in the tree. 
 The \emph{Markov chain tree theorem} (see \citet{FreWen84}, Chapter~6, Lemma 3.1) says that, for any ergodic Markov chain with transition matrix $P$, its stationary distribution $\pi$ is given by: 
 \begin{align}\label{eq:tree-stationary}
 	\pi(s) =& \frac{W(s)}{\sum_{\hat{s}}W(\hat{s})} \enspace, 
 	& \text{where }& & W(s) :=& \sum_{{T\in \mathcal T(s)}} \P(T) \enspace . 
 \end{align}

\subsection{Measure of Disorder}
In this section we introduce a formal definition for the \textit{total weighted inversion}, which can be seen as a measure of disorder. As we shall prove in the next section, this arises naturally from the algorithm performing adjacent swaps.

	\begin{definition} 
		The \textit{total weighted inversion} of a sequence $s$ is defined as $$w(s) := \sum_{i<j \colon s_i>s_j}{s_i-s_j}\enspace .$$
	\end{definition}
	
	\begin{example}
		Consider the sequence $s=(5,2,3)$ and the sorted sequence $(2,3,5)$. Then the total weighted inversion of $s$ is equal to $w(s)=(5-2)+(5-3) = 5$.
	\end{example}
	
	The displacements of the single elements allow an equivalent way to describe the total weighted inversion (this equivalent definition turns out to be useful in the next section).
	
	\begin{lemma}
		For a sequence $s$ let $s^{(sort)}$ be the sequence sorted in non-decreasing order. Then, $w(s) = \sum_i (s^{(sort)}_i - s_i)i$.
	\end{lemma}
	\begin{proof}
		In the sum $\sum_{i<j \colon s_i>s_j}{s_i-s_j}$, every element $s_i$ is added $r_i$ and subtracted $l_i$ times, where $r_i$ is the number of smaller elements on its right hand side and $l_i$ the number of larger elements on its left. The difference $d_i = r_i -l_i$ corresponds to the diplacement of $s_i$ to the right compared to the sorted sequence, i.e., $s^{(sort)}_{i+d_i} = s_i$. Since $d_i \cdot s_i$ is exactly the contribution of $s_i$ to $w(s)$, the claim follows immediately.
	\end{proof}

\section{Sorting algorithms as Markov chains}
In this section we define the algorithms and the resulting Markov chains. The first chain performs only adjacent comparisons. 

\begin{definition}
	The chain $\MCadj$ is defined as follows: 
	\begin{enumerate}  
		\item Pick an index $i$ in $\{1,\ldots,n-1\}$ uniformly at random;
		\item Swap $s_i=a$ and $s_{i+1}=b$ with probability $\frac{\lambda^{a-b}}{\lambda^{a-b}+\lambda^{b-a}}$.
	\end{enumerate}
\end{definition}  

We shall prove below that the stationary distribution of this chain assigns higher probabilities to the sequences that are ``nearly sorted''.

\begin{theorem}
	The chain $\MCadj$ is reversible with stationary distribution 
	$\pi(s) \propto \lambda^{-2w(s)}$,
	where $w(s)$ is the total weighted inversion.
\end{theorem}
\begin{proof}
	We  prove that the chain is reversible. Let $s$ and $s'$ be two sequences that differ in $i$'s swap (otherwise $P(s,s')=0=P(s',s)$ and reversibility is trivial). Observe that by definition 
	\begin{align*}
	\frac{P(s,s')}{P(s',s)} =& \lambda^{2(s_{i} - s_{i+1})}=\lambda^{2(a-b)} & \text{ and} &  & \frac{\pi(s')}{\pi(s)} = \lambda^{2w(s) - 2w(s')} \enspace .
	\end{align*}
	Since $s^{(sort)}=s'^{(sort)}$ and  $s'$ is obtained by swapping $a=s_i$ and $b=s_{i+1}$, 
	\begin{align*}
		w(s) - w(s') = \enspace\enspace\enspace & (s^{(sort)}_i- s_i)i -  (s^{(sort)}_i-s'_i)i &\\
		+   & (s^{(sort)}_{i+1}-s_{i+1})(i+1) - (s^{(sort)}_{i+1}-s'_{i+1})(i+1) 
		& = a- b 
	\end{align*}
	and therefore the detailed balance condition is satisfied.
\end{proof}

We next consider chains which compare any two numbers in the sequence:

\begin{definition}
	The chain $\MCany$ is defined as follows: 
	\begin{enumerate}
		\item Pick two indexes $i$ and $j$ in $\{1,\ldots,n\}$ uniformly at random, with $i< j$;
		\item Swap $s_i=a$ and $s_j=b$ with probability $\frac{\lambda^{a-b}}{\lambda^{a-b}+\lambda^{b-a}}$. 
	\end{enumerate}
	The chain $\MCanyrev$ is defined as above except that the probability of swapping is $\left(\frac{\lambda^{a-b}}{\lambda^{a-b}+\lambda^{b-a}}\right)^{j-i}$.
\end{definition}

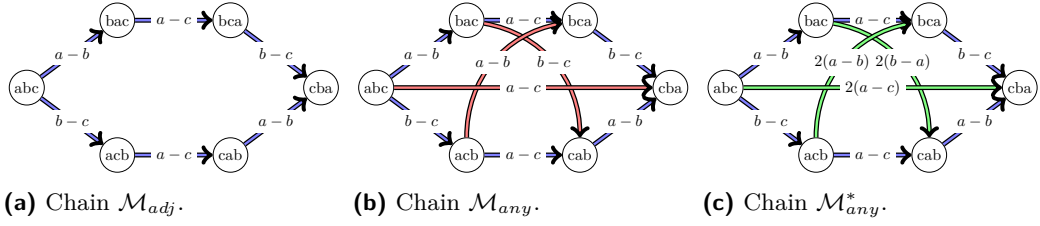
\begin{figure}
	\centering
	\begin{subfigure}[b]{0.32\textwidth}
		\centering
		\resizebox{\linewidth}{!}{
			\begin{tikzpicture}
			\states
			\abcbac \abcacb \bacbca \acbcab \bcacba \cabcba
			\end{tikzpicture}
		}
		\caption{Chain $\MCadj$.}
		\label{fig:MCadj}
	\end{subfigure}
	\begin{subfigure}[b]{0.32\textwidth}
		\centering	
		\resizebox{\linewidth}{!}{	
			\begin{tikzpicture}
			\states
			\abcbac \abcacb \bacbca \acbcab \bcacba \cabcba
			\edgeMany
			\baccab \acbbca \abccba
			\end{tikzpicture}
	}	
		\caption{Chain $\MCany$.}
		\label{fig:fig:MCany}
	\end{subfigure}		
	\begin{subfigure}[b]{0.32\textwidth}
		\centering		
		\resizebox{\linewidth}{!}{
			\begin{tikzpicture}
			\states
			
			\abcbac \abcacb \bacbca \acbcab \bcacba \cabcba
			\edgeManyrev
			\Edge[label=$2(b-a)$, style={bend left = 40}](bac)(cab);
			\Edge[label=$2(a-b)$, style={bend left = 40}](acb)(bca);
			\Edge[label=$2(a-c)$](abc)(cba);
			\end{tikzpicture}	
		}
		\caption{Chain $\MCanyrev$.}
		\label{fig:fig:MCany-rev}
	\end{subfigure}
	
	\label{fig:chains:abc}
	\caption{The three chains for sorting three elements $abc$; A transition with label $w$ has  probability  $\frac{\lambda^w}{\lambda^w + \lambda^{-w}}$; For clarity sake we only show forward transitions.}
\end{figure}

\begin{theorem}\label{th:anyrev:stationary}
	Chain $\MCanyrev$ is reversible and has the same stationary distribution as $\MCadj$, while chain $\MCany$ is not reversible. 
\end{theorem}
\begin{proof}
	\newcommand{\MCpath}{\mathcal{P}}
	To prove that $\MCanyrev$ has the same stationary distribution as $\MCadj$ we argue as follows. 
	Consider any transition from $s$ to $s'$ which swaps two elements at distance $k\geq 1$. There exists a path $\MCpath$ in $\MCadj$ that leads from $s$ to $s'$ and whose probability has the same form of the single transition in $\MCanyrev$,
	\begin{align*}
	\P(\MCpath)=\prod_{(x,y) \in \MCpath} P(x,y) = \frac{\lambda^{k(a-b)}}{D(\MCpath)} \enspace .
	\end{align*}
	The path is obtained by simulating the swap between $a$ and $b$ via adjacent swaps:
	{\small \begin{align*}
	\MCpath := &(\cdots a x_1 \cdots x_{k-1} b \cdots) \leftarrow (\cdots  x_1 a \cdots x_{k-1} b \cdots) \cdots \leftarrow 
	 (\cdots   x_1 \cdots x_{k-1} a b \cdots) \leftarrow\\ &(\cdots   x_1 \cdots x_{k-1} ba \cdots) \leftarrow  
	 (\cdots x_1 x_2 \cdots b x_{k-1}a \cdots) 
	 \cdots 
	\leftarrow (\cdots b x_1 x_2 \cdots x_{k-1} a \cdots)
	\end{align*} }
	which yields in the numerator the product
	\[
	\lambda^{(a-x_1)+\cdots+ (a-x_{k-1}) + (a-b) + (x_{k-1}-b) + \cdots +(x_1-b)} = \lambda^{k(a-b)} \enspace .
	\]
	Note also that the reverse path $\MCpath^{-1}$ leading from $s'$ to $s$ has probability
	$
	\P(\MCpath^{-1}) = \frac{\lambda^{k(b-a)}}{D(\MCpath)}\enspace,
	$
	where the denominator $D(\MCpath)$ is the same as above because  all transitions in the chains are of the form $P(x,y)=N_{xy}/D_{xy}$ with $D_{xy}=D_{yx}$. Since $\MCadj$ is reversible, we get the first of the following equalities:
	\[
	\frac{\pi(s')}{\pi(s)} = \frac{\P(\MCpath)}{\P(\MCpath^{-1})} = \frac{\lambda^{k(a-b)}}{\lambda^{k(b-a)}} =  \frac{P^*(s,s')}{P^*(s',s)}\enspace,
	\]
	where $P^*$ is the transition matrix of $\MCanyrev$. 
	Thus, the detailed balance condition for $P^*$ is satisfied and $\pi$ is the stationary distribution of $\MCanyrev$. 
	
Finally, to see that $\MCany$ is not reversible, consider the the cycle $(abc)\leftarrow (bac) \leftarrow (bca) \leftarrow (cba) \leftarrow (abc)$ in Figure~\ref{fig:fig:MCany} which has a length different from the reversed cycle. This violates the  Kolmogorov reversibility criterion \eqref{eq:Kolmogorov_criterion}. 
\end{proof}

In our experiments (see Section~\ref{sec:experiments}), it turns out that doing only adjacent comparisons ($\MCadj$) is better than doing any comparisons ($\MCany$);  The following  sections provide  analytical results for special cases. 
Note that Theorem~\ref{th:anyrev:stationary} says that in the long-run $\MCanyrev$ is as good as $\MCadj$.

\section{Binary inputs}\label{sec:binary}
In this section we restrict to the case in which every  element in the sequence is either $a$ or $b$ for some $b>a$. That is, the sorted sequence is $(a,\ldots,a,b,\ldots,b)$, where $n_a$ denotes the number of $a$'s and $n_b$ denotes the number of $b$'s.

\subsection{Mixing time}
For binary inputs, we can uniquely express every sequence by a vector $v\in \{0,\dots,n_a\}^{n_b}$, $v_1\geq v_2\geq \dots\geq v_{n_b}$, where $v_i$ denotes the number of inversions of the $i$-th $b$  in the sequence (for example, $babba$ corresponds to $211$, while $babab$ corresponds to $210$).
Such a vector is visualized as a monotonically decreasing `staircase' in a $n_b \times n_a$ grid and $\MCadj$ corresponds to the biased Markov process in \cite{GreenbergPR09}. 
The bounds on the mixing time for this process translate immediately for our chain.

\begin{theorem}[by Theorem~2.1 in \cite{GreenbergPR09}]
	For binary inputs, the mixing time of $\MCadj$ satisfies \[t_{mix}(\epsilon) = O(n^2 \log (\epsilon^{-1})).\]
\end{theorem}
Observe also that the chain $\MCadj$ corresponds to the well-known \emph{asymmetric simple exclusion process} (see \cite{Benetal05} and \cite{Bhaetal13}). 

We next consider $\MCany$ and prove an upper bound. To bound the mixing time of $\MCany$ we use the method of \emph{path coupling} \cite{DyerG98}. A \emph{path coupling} for a
chain $\mathcal M$ can be specified by providing distributions
\begin{align}
\mathbb{P}_{x,y}&[X=x', \, Y=y'], & \text{ for all }
&x,y\in S \text{ such that } P(x,y)>0, \label{eq:coupling:move}
\intertext{satisfying, for all $x,y\in S$ such that
	$P(x,y)>0$,} 
\mathbb{P}_{x,y}&[X=x']= P(x,x')
& \text{ for all }&x' \in S, \label{eq:coupling-x}\\  
\mathbb{P}_{x,y}&[Y=y']= P(y,y')
& \text{ for all }&y' \in S. \label{eq:coupling-y}  
\end{align}
We use $\rho$ to denote the shortest-path distance in the Markov
chain, i.e., $\rho(x,y)$ is the minimum number of transitions to go
from $x$ to $y$.

\begin{lemma}[Theorem~2.1 in \cite{DyerG98}]\label[lemma]{thm:mixingtime}\label{th:path-coupling}
	Suppose there exists $\beta<1$ such that, for all $x,y$ with $P(x,y)>0$, it holds that 
	\begin{equation}
	\label{eq:path-coupling:contraction}
	\mathbb E_{x,y}[\rho(X,Y)]\leq \beta.
	\end{equation} Then the mixing time $t_{mix}(\epsilon)$ of the Markov chain under consideration satisfies \[t_{mix}(\epsilon)\leq \frac{\log(D \epsilon^{-1})}{1 - \beta}.\] 
\end{lemma}

\paragraph*{Path coupling for $\MCany$.}

Consider two sequences $x$ and $y$ which differ by swapping elements in position $i^*$ and $j^*$.
For every such pair $(x,y)$ we specify the probabilities in \eqref{eq:coupling:move} to move to a pair $(x',y')$.
We group the ${n \choose 2}$ different swaps between elements in positions $i$ and $j$ as follows:
\begin{align*}
(i^*,j^*) \leftrightarrow (i^*,j^*) &&
(i^*,j) \leftrightarrow (j^*,j) &&
(i,j^*) \leftrightarrow (i,i^*) &&
(i,j) \leftrightarrow (i,j),
\end{align*}
in the sense that if we consider the positions $i^*$ and $j$ in one sequence, we consider the positions $j^*$ and $j$ in the other sequence, and vice versa.
Clearly, this defines a bijection on the swaps of the two sequences.
Now let $x_\mathsf{i\times j}$ denote the sequence obtained from $x$
by swapping the two elements at positions $i$ and $j$.
The path coupling is as follows:
\begin{flalign*}
&\text{for $i^*,j^*$} &
(x,y)&\mapsto (y,y) & \mbox{with \enspace}
&  P(x,y), 
\\
&&(x,y)&\mapsto (x,x)  & \mbox{with \enspace}
& P(y,x), 
\\
&\text{for $i^*,j$} &
(x,y) &\mapsto (x_\mathsf{i^*\times j},y_\mathsf{j^*\times j})
&\mbox{with \enspace} &
\min\{P(x,x_\mathsf{i^*\times j}),P(y,y_\mathsf{j^*\times j})\},
\\
&&(x,y) &\mapsto (x_\mathsf{i^*\times j},y) & \mbox{with \enspace}
& \max\{0, P(x,x_\mathsf{i^*\times j})-P(y,y_\mathsf{j^*\times j})\},
\\
&&(x,y) &\mapsto (x,y_\mathsf{j^*\times j}) & \mbox{with \enspace}
& \max\{0, P(y,y_\mathsf{j^*\times j})-P(x,x_\mathsf{i^*\times j})\},
\\  
&\text{for $i,j^*$} &
(x,y) &\mapsto (x_\mathsf{i\times j^*},y_\mathsf{i\times i^*})
&\mbox{with \enspace} &
\min\{P(x,x_\mathsf{i\times j^*}),P(y,y_\mathsf{i\times i^*})\},
\\
&&(x,y) &\mapsto (x_\mathsf{i\times j^*},y) & \mbox{with \enspace}
& \max\{0, P(x,x_\mathsf{i\times j^*}) -P(y,y_\mathsf{i\times i^*})\},
\\
&&(x,y) &\mapsto (x,y_\mathsf{i\times i^*}) & \mbox{with \enspace}
& \max\{0, P(y,y_\mathsf{i\times i^*})-P(x,x_\mathsf{i\times j^*})\},
\\  
&\text{for $i,j$} &
(x,y) &\mapsto (x_\mathsf{i\times j},y_\mathsf{i\times j})
&\mbox{with \enspace} &
P(x,x_\mathsf{i\times j})=P(y,y_\mathsf{i\times j}).
\intertext{Finally, with all remaining probability} &&
(x,y) &\mapsto (x,y). & & 
\end{flalign*}
One can easily check that this is indeed a path coupling, that is,
\eqref{eq:coupling-x}-\eqref{eq:coupling-y} are satisfied. The
difficulty is in proving the condition necessary to apply
Lemma~\ref{th:path-coupling}.

\begin{lemma}\label[lemma]{thm:coupling-grid}
	The path coupling defined above satisfies condition
	\eqref{eq:path-coupling:contraction} with \[\beta \leq 1 -\frac{2(1+p(n-2))}{n(n-1)}  \leq 1 -\frac{2p}{n-1}.\]
\end{lemma}
\begin{proof}
	The second inequality follows from $p(n-2) + 1 \geq pn$, since $p \leq \frac{1}{2}$. We next prove the first inequality. 
	Let $x,y$ be two sequences that differ in swapping positions $i^*$ and $j^*$, thus $\rho(x,y) = 1$.
	Since $P(x,y) + P(y,x) = 1$, the new distance $\rho(x',y')$ after choosing positions $i^*$ and $j^*$ is always zero.
	Furthermore, for every position $k$, such that $k\neq i^*$ and $k\neq j^*$, either $\rho(x_\mathsf{i^*\times k},y_\mathsf{j^*\times k})=0$ or $\rho(x_\mathsf{k\times j^*},y_\mathsf{k\times i^*})=0$, and the probability of accepting such a transition is at least $p$.
	Finally, it is easy to see that after every other transition $\rho(x',y') = 1$.
	Remember that there are ${n \choose 2}$ pairs of positions in a sequence.
	Therefore, $\mathbb{E}[\rho(x',y')] \leq \left(1 - \frac{1 + p(n-2)}{{n \choose 2}}\right)$.
\end{proof}

\begin{theorem} Let $n_a$ (resp., $n_b$) denote the number of $a$'s (resp., $b$'s) in the sequence.
	The mixing time of $\MCany$ satisfies  \[t_{mix}(\epsilon)\leq \frac{n(\log(n') - \log(\epsilon))}{2p},\] where $n'=\min\{n_a,n_b\}\leq \frac{n}{2}$.
\end{theorem}
\begin{proof}
	The diameter $D$ is the maximum number of transitions required to go from any sequence to any other sequence. 
	Then $D = \min\{n_a,n_b\} \leq \frac{n}{2}$, because with $D$ swaps we can either move all $a$ or all $b$ to their desired positions. By  \Cref{thm:mixingtime,thm:coupling-grid}, the claim follows immediately.
\end{proof}

\section{Only adjacent swaps is better}\label{sec:stationary-compare}
In this section we prove that, for two special cases, the chain $\MCadj$ performing only adjacent comparisons is better than the chain $\MCany$ performing comparisons between any two elements. 

\subsection{Three elements}   
Our first special case is to consider sorting three arbitrary elements and show that $\MCadj$ has more chances to return the sorted sequence than $\MCany$. 

\begin{theorem}\label{thm:adj-is-better}
	For any three elements, not all of them identical, the chain $\MCadj$ returns the sorted sequence with a probability (at stationary distribution) strictly larger than that of $\MCany$ (at stationary distribution), 
	\begin{equation}\label{eq:MCany-worse}
		\pi_{adj}(abc) > \pi_{any}(abc)
	\end{equation}
	where $abc$ is the sorted sequence ($a \leq b\leq c$), for all $\lambda>0$.
\end{theorem}

In order to prove this theorem we show that the 
ratios between the distribution of adjacent states in $\MCadj$ gets ``worse'' in $\MCany$:

\begin{restatable}{lemma}{ratiosabc}\label{le:ratios-abc}
	For any two states $s$ and $s'$ that differ in exactly one adjacent swap, 
	if the total weighted inversion satisfies $w(s') > w(s)$, then it holds that
	\begin{equation}\label{eq:lemma1}
	\frac{\pi_{adj}(s')}{\pi_{adj}(s)} < \frac{\pi_{any}(s')}{\pi_{any}(s)} \enspace .
	\end{equation}
\end{restatable}

\begin{proof}[Proof Idea]
We  use the Markov Chain Tree Theorem \eqref{eq:tree-stationary}. Our goal is thus to show that 
\begin{align}\label{eq:bound:trees-ratio}
	\frac{\sum_{T\in \mathcal T(s) } \P(T)}{\sum_{T'\in \mathcal T(s')} \P(T)} < \frac{\pi_{adj}(s)}{\pi_{adj}(s')} = \lambda^{2(w(s')-w(s))} && \text{for} && w(s) < w(s') \enspace .
\end{align}
Ideally, one would like to find a bijection from trees $T\in \mathcal{T}(s)$ to trees $T'\in \mathcal{T}(s')$ such that $\frac{\P(T)}{\P(T')} < \frac{\pi_{adj}(s)}{\pi_{adj}(s')}$ holds for each tree $T\in \mathcal{T}(s)$. Unfortunately, this is in general not possible, so the following slightly more involved argument is used: 
\begin{itemize}
	\item The simple mapping we use consists in reversing the path from $s'$ to $s$ in $T$ to obtain the new tree $T'$. This mapping is a bijection between  $\mathcal T(s)$ and $\mathcal T(s')$.
	\item Because this mapping does not guarantee the desired inequality for all trees $T$, we classify the trees in $\mathcal{T}(s)$ into \emph{good} and \emph{bad} trees: a tree $T$ is \emph{bad} if $\frac{\P(T)}{\P(T')}> \lambda^{2(w(s')-w(s))}$ and \emph{good} otherwise. We then show that
	\begin{align*}
		\frac{\sum_{T\in bad}\P(T) + \sum_{T\in good}\P(T)}{\sum_{T\in bad}\P( T') + \sum_{T\in good}\P( T')} < \lambda^{2(w(s')-w(s))} \enspace ,
	\end{align*}
	where $T'$ is the tree obtained from $T$ via the mapping in the previous item. This proves \eqref{eq:bound:trees-ratio} since $good$ and $bad$ define a partition of $\mathcal T(s)$ and also a partition of $\mathcal T(s')$.
\end{itemize}
The details of this proof are given in Appendix~\ref{app:omitted-proofs}.
\end{proof}

From this lemma it is easy to obtain \eqref{eq:MCany-worse} in \Cref{thm:adj-is-better}.
\begin{proof}[Proof of \Cref{thm:adj-is-better}]
	By transitivity,  Lemma~\ref{le:ratios-abc} implies that, for all non-sorted sequences $s \neq (abc)$,
	$
	\frac{\pi_{adj}(s)}{\pi_{adj}(abc)} < \frac{\pi_{any}(s)}{\pi_{any}(abc)},
	$
	and therefore 
	$$
	\pi_{adj}(abc) =\frac{1}{1 + \sum_{s \neq (abc)} \frac{\pi_{adj}(s)}{\pi_{adj}(abc)}} > \frac{1}{1 + \sum_{s \neq (abc)} \frac{\pi_{any}(s)}{\pi_{any}(abc)}} = \pi_{any}(abc).
	$$
\end{proof}

	\subsection{One outlier}\label{sec:equals}
	We call \emph{one outlier} the case in which we have $n-1$ small identical elements, and only one bigger element (the outlier) to be sorted. That is, the sorted sequence is  $(a,a,\dots,a,b)$, with $b>a$. Since swapping two identical elements does not change the sequence, i.e. the state of the chain, we have $n$ states which correspond to the possible positions of $b$. We denote the state in which $b$ is in position $i$ by $s^{(i)}$, so that $s^{(n)}=(a,\dots,a,b)$ is the sorted sequence and  $s^{(1)} := (b,a,\dots,a)$ is the ``reversely sorted'' sequence. Note that the weighted inversion of $s^{(i)}$ is $(n-i)(b-a)$, thus implying that the expected total weighted inversion is of the form
	\begin{equation}\label{eq:expected_diclos:generic}
	E^{w} := \sum_{i=1}^n (n-i)(b-a)\pi(s^{(i)}) \enspace .
	\end{equation}
	It is useful for the analysis to consider the probability that element $b$ is erroneously declared smaller than $a$ in a single comparison, 
	\[
	p :=1 - p_{ab}=\frac{\lambda^{a-b}}{\lambda^{a-b}+\lambda^{b-a}}.
	\]
	\begin{observation}
For the one outlier case, the chain $\MCadj$ becomes a so-called \textit{birth-and-death} chain  meaning that from each state $s^{(i)}$ we can only move to $s^{(i+1)}$ or to  $s^{(i-1)}$, and the transition probabilities are
\begin{align*}
	P(s^{(i+1)},s^{(i)})&=\frac{p}{n-1} \ , && P(s^{(i)},s^{(i+1)}) = \frac{1-p}{n-1} \ , 
\end{align*}
for all $ i\in \{1,\dots,n-1\}$. In the chain $\MCany$ every state is connected to all other states and the transition probabilities are $P(s^{(i)},s^{(j)})= p/{n \choose 2}$ if $i>j$ and $(1-p)/{n \choose 2}$ if $i < j$.	
	\end{observation}

The next theorem says that the chain $\MCadj$ has a better probability of returning the sorted sequence and a better expected total weighted inversion than  $\MCany$. 
\begin{theorem}\label{th:outlier}
	For the case of one outlier,  the following holds.
	The probability of obtaining the sorted sequence, at stationary distribution, is constant for the $\MCadj$, while for $\MCany$ it converges to zero as $n$ grows:
			\begin{align*}
				\pi_{adj}(s^{(sorted)}) &> \frac{1-2p}{1-p}\enspace , & 
				\pi_{any}(s^{(sorted)}) &< \frac{1-p}{np}\enspace.
				\intertext{The expected total weighted inversion is constant for $\MCadj$, while for $\MCany$ it grows linearly in $n$:}
		E^{w}_{adj} & < \frac{p}{1-2p}(b-a)\enspace, &
		E^{w}_{any} & > np (b-a)\enspace.
	\end{align*}		
\end{theorem}
The theorem above follows immediately from the next two lemmas, whose proof is given in Appendix~\ref{app:simple-chain}. Recall that $s^{(sorted)})=s^{(n)}$.
	
	\begin{restatable}{lemma}{statdist}\label{lem:stat-prob}	
		The stationary distributions of $\MCadj$ and $\MCany$, are 
		\begin{align*}
			\pi_{adj}(s^{(i)})&= \frac{p^{n-i}(1-p)^{i-1}(1-2p)}{(1-p)^n-p^n}\enspace ,\\
			\pi_{any}(s^{(i)}) &= \frac{np(1-p)}{((n-i+1)(1-p)+(i-1)p)((n-i)(1-p)+ip)}\enspace.
		\end{align*}		
	\end{restatable}
	
	\begin{restatable}{lemma}{expweight}\label{lem:exp-disl}
		For $\MCadj$ and $\MCany$, the corresponding expected total weighted inversions  are
		\begin{align*}
			E^{w}_{adj} 
			&= n(b-a)p\left(\frac{1}{n(1-2p)}-\frac{p^{n-1}}{(1-p)^n-p^n}\right) < (b-a)\frac{p}{1-2p}\enspace, \\
			E^{w}_{any} 
			&= n(b-a)p\cdot \sum_{i=0}^{n-1} \frac{i(1-p)}{((i+1)(1-p)+(n-i-1)p)(i(1-p)+(n-i)p)}> n(b-a)p \enspace.
		\end{align*}
	\end{restatable}

\section{Experimental results}\label{sec:experiments}
\begin{figure}[h!]
	\minipage{0.48\textwidth}
	\includegraphics[width=\linewidth]{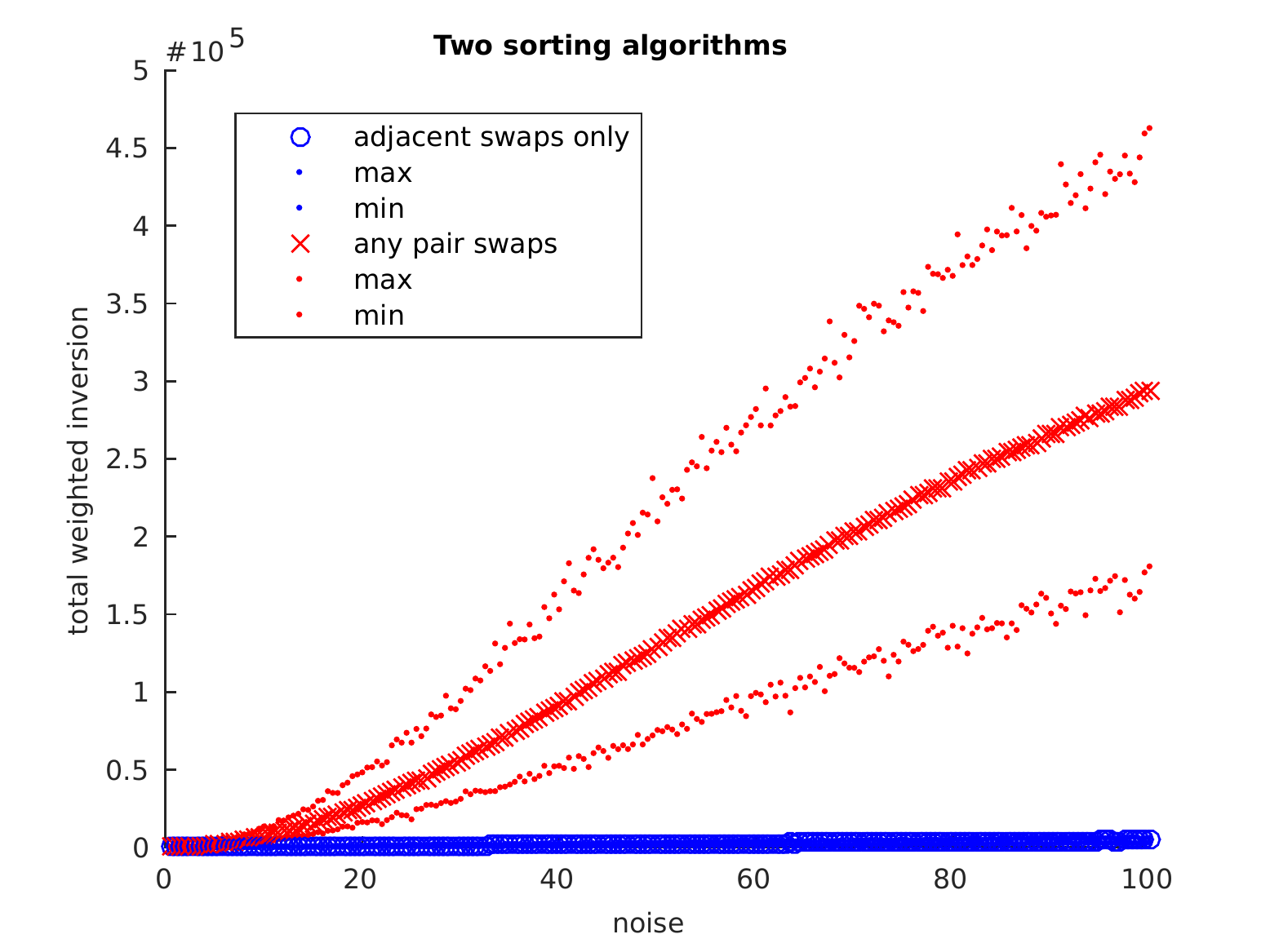}
	\endminipage\hfill
	\minipage{0.48\textwidth}
	\includegraphics[width=\linewidth]{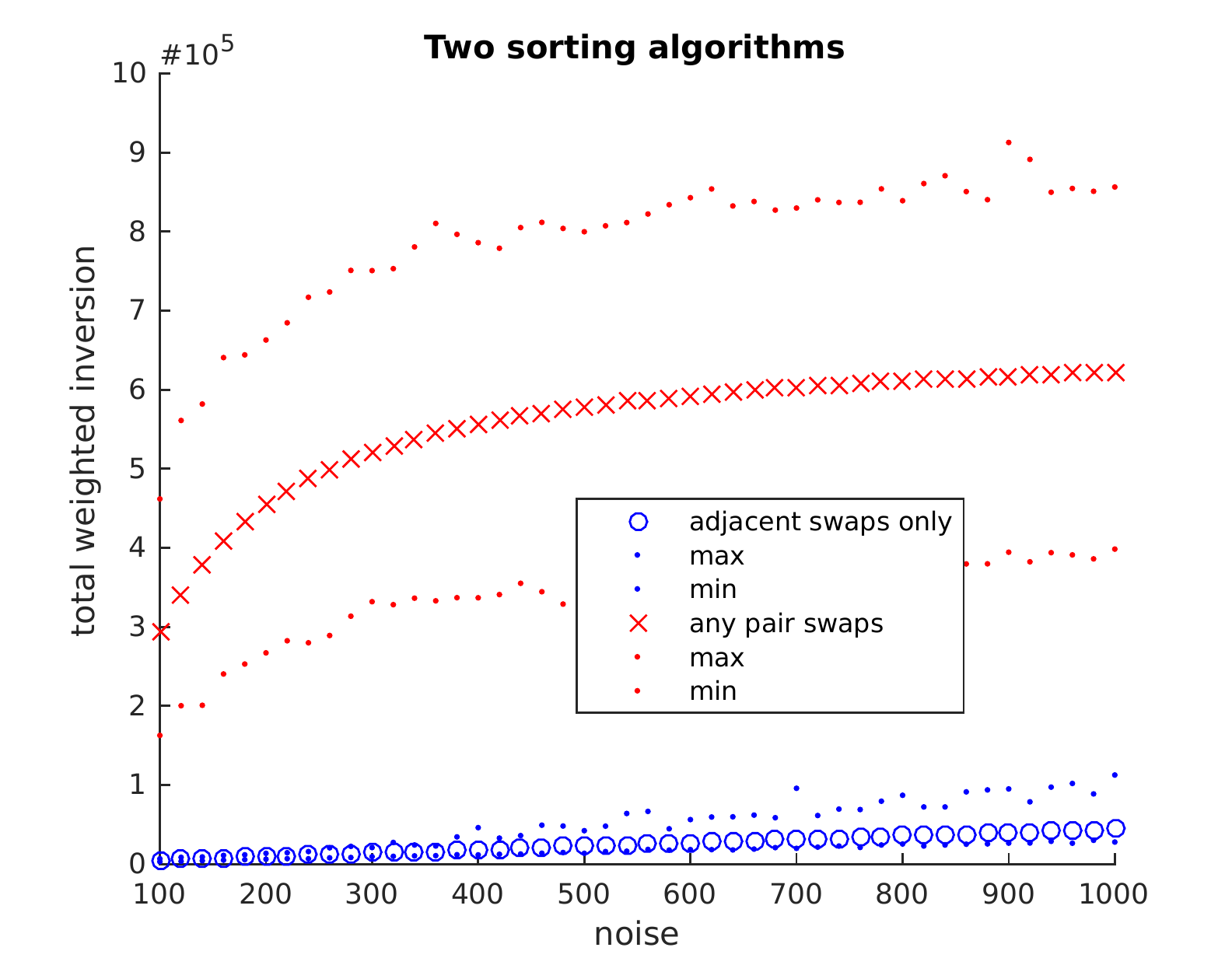}
	\endminipage\hfill
	
	\minipage{0.48\textwidth}%
	\includegraphics[width=\linewidth]{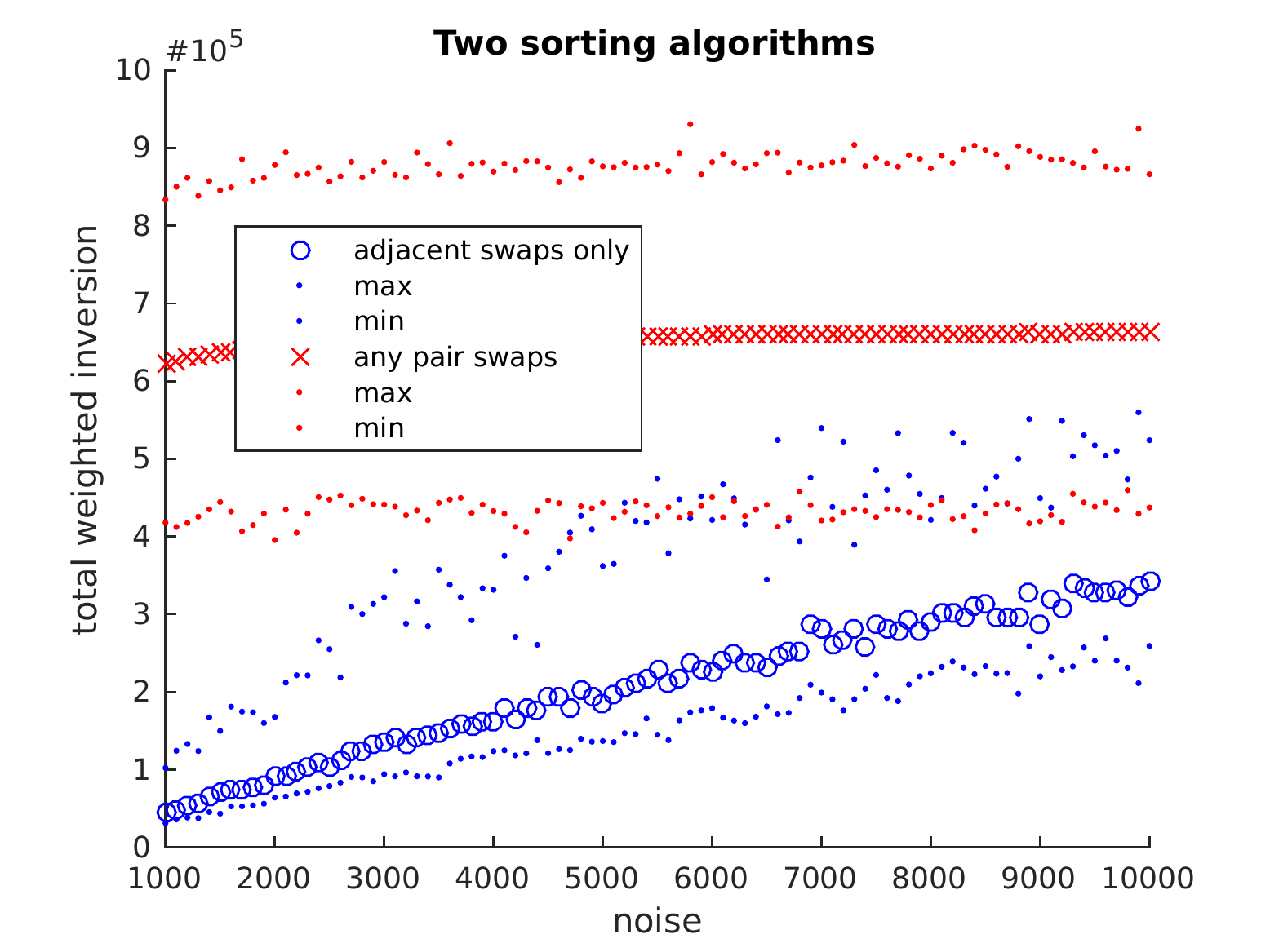}
	\endminipage\hfill
	\minipage{0.47\textwidth}%
	\caption{A comparison of $\MCadj$ and $\MCany$ on the long run. 
	We measure the average, maximum, and minimum total weighted inversion for a certain number of steps after both algorithms have approached their stationary distribution. The elements to be sorted are $(200,199,\ldots,1)$.  By increasing the value of $noise$, where  $\lambda=e^{1/noise}$,
	the stationary behavior of $\MCany$ degrades much earlier than that of $\MCadj$.
	}\label{fig:exp-weighted-disl}
	\endminipage	
\end{figure}

\begin{figure}[h!]
	\minipage{0.48\textwidth}
	\includegraphics[width=\linewidth]{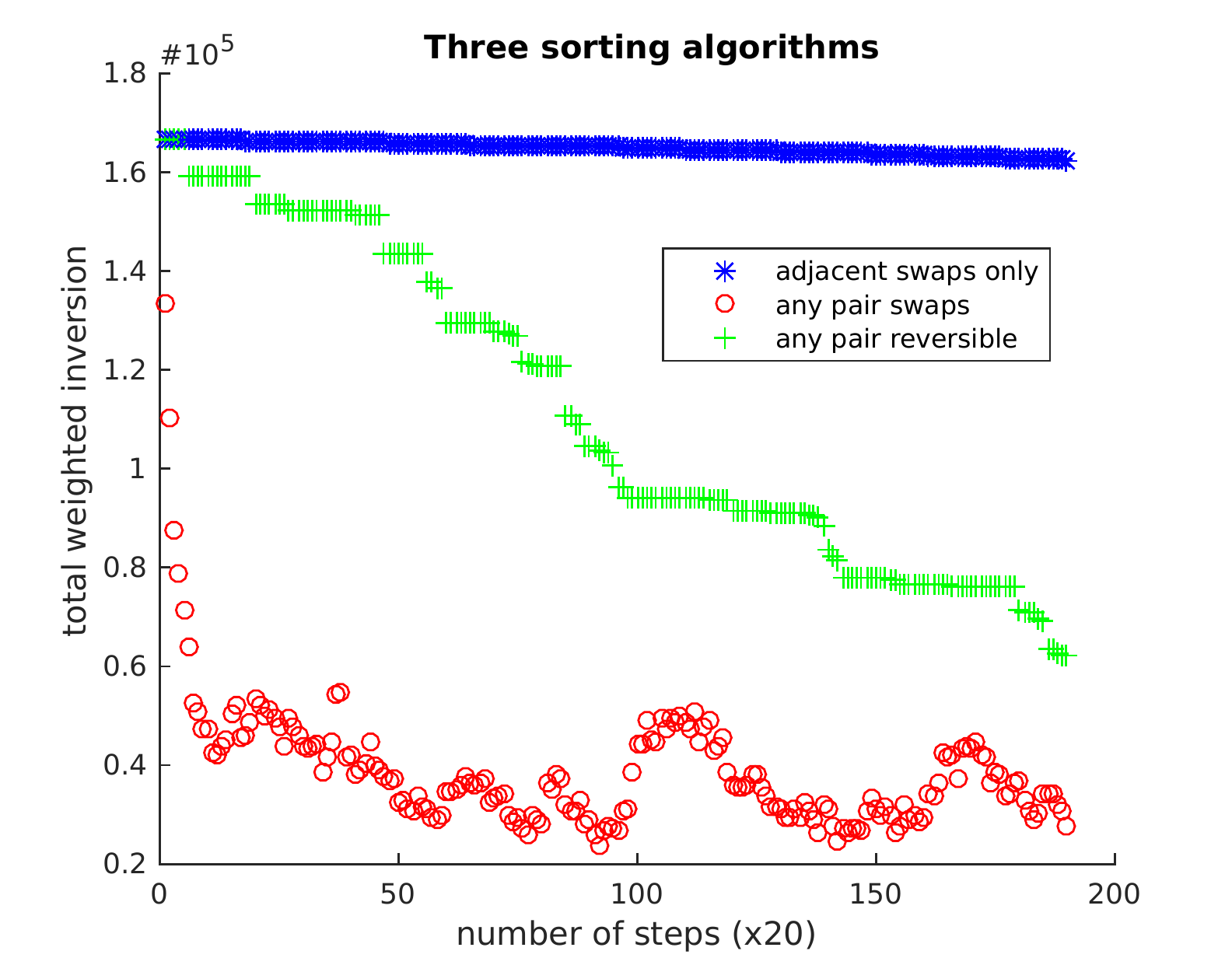}
	\endminipage\hfill
	\minipage{0.48\textwidth}
	\includegraphics[width=\linewidth]{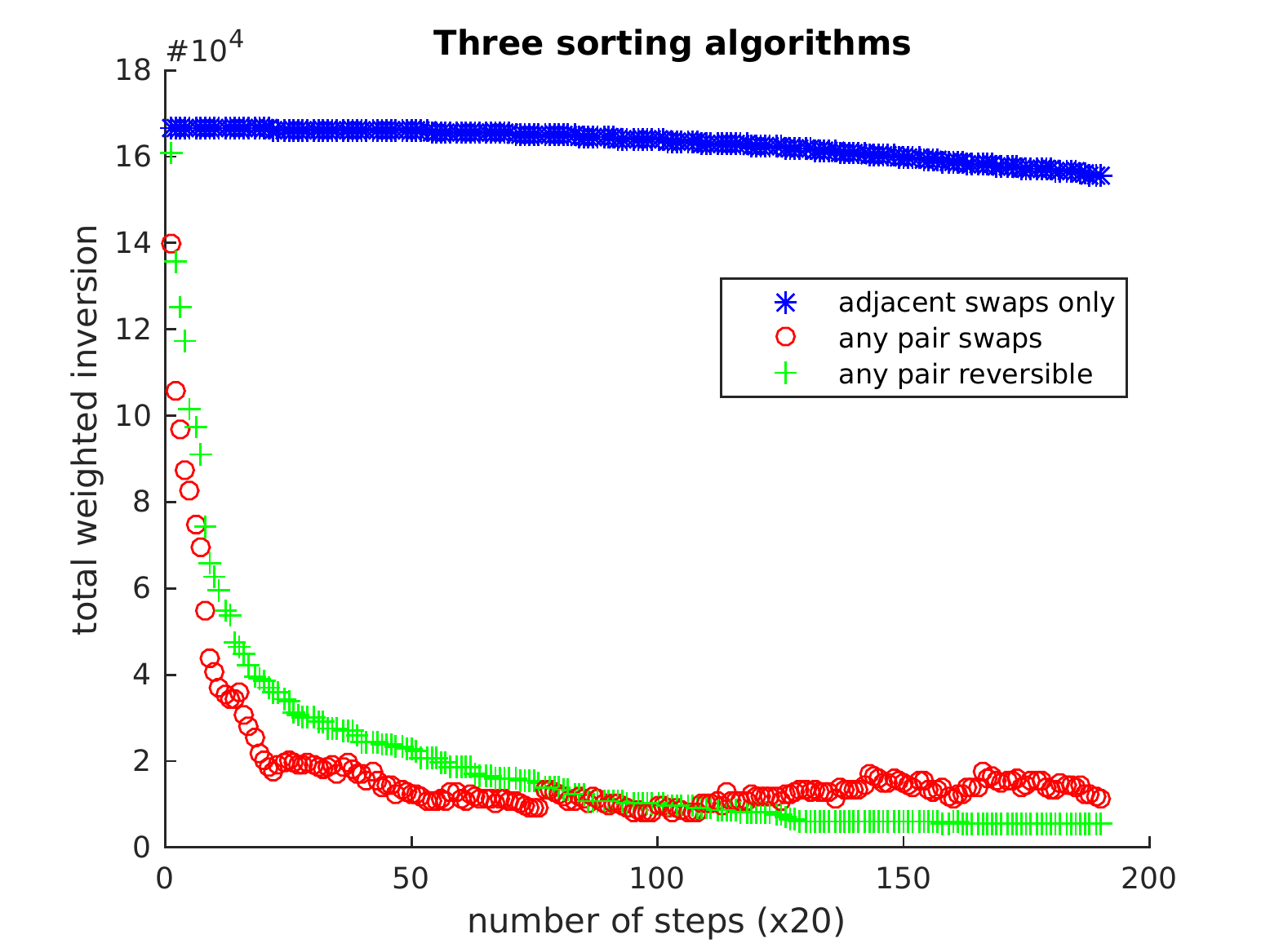}
	\endminipage\hfill
	
	\minipage{0.48\textwidth}%
	\includegraphics[width=\linewidth]{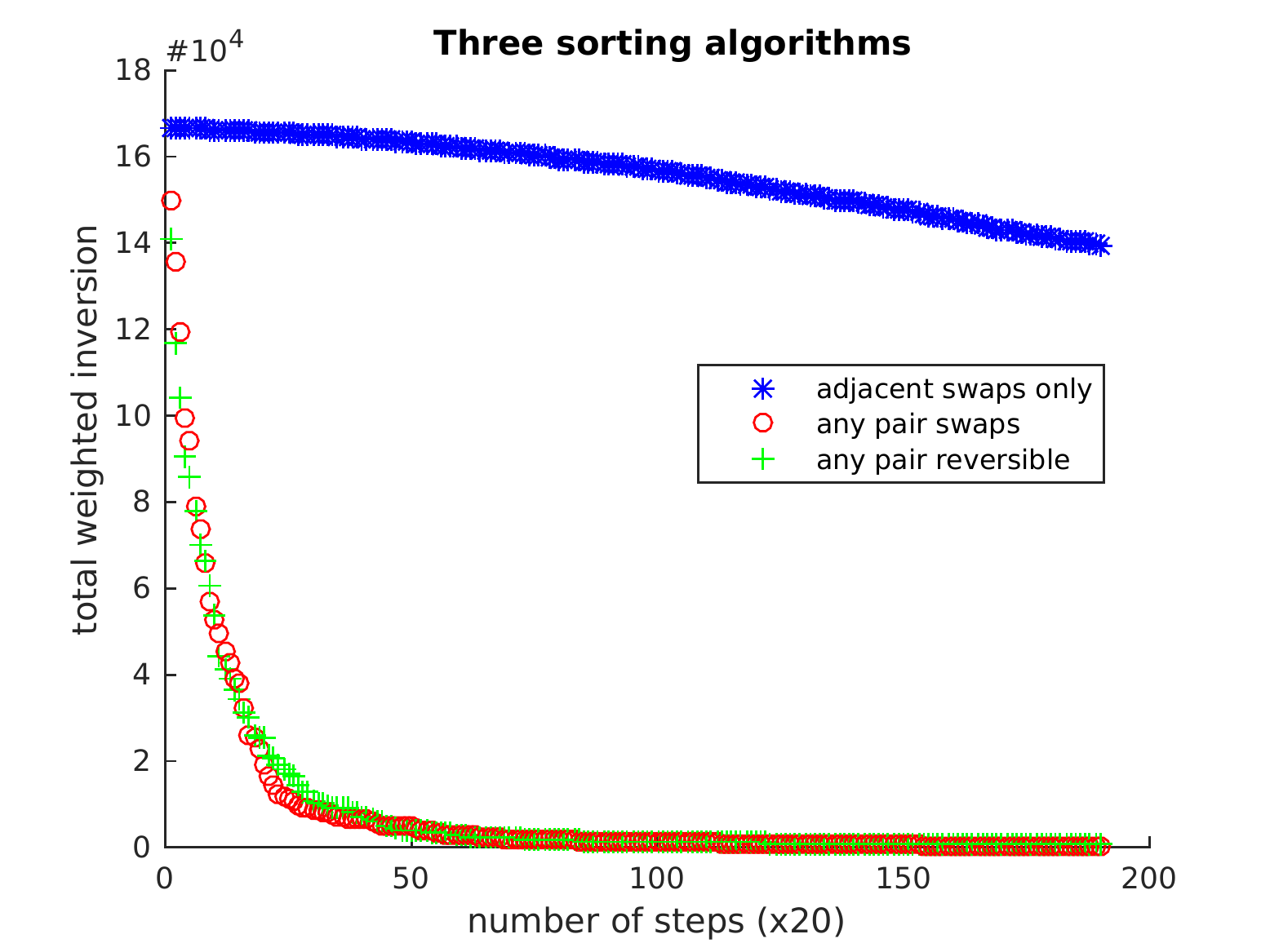}
	\endminipage\hfill
	\minipage{0.47\textwidth}%
	\caption{A comparison of the three algorithms $\MCadj$, $\MCany$, and $\MCanyrev$. The elements to be sorted are $(100,99,\dots,1)$. The plots are for three different values of noise $\{50,20,0.1\}$, where $\lambda=e^{1/noise}$.
	}\label{fig:expereminets:all_three}
	\endminipage		
\end{figure}

We conducted the following set  of experiments on several input sequences to compare the two sorting algorithms $\MCadj$ and $\MCany$. 
\begin{itemize}
	\item \textbf{Low energy (high noise) regime.} We evaluate how much the two algorithms are robust to an increase of the error probability by taking $\lambda=e^{1/noise}$ for increasing values of $noise$. Figure~\ref{fig:exp-weighted-disl} shows that $\MCany$ degrades much earlier than $\MCany$. 
	\item \textbf{The best of the two.} We compare all three algorithms $\MCadj$, $\MCany$ and $\MCanyrev$ in Figure~\ref{fig:expereminets:all_three}. These experiments suggest that $\MCanyrev$ possesses good features from both the other algorithms: the total weighted inversion decreases faster than $\MCadj$, while its stationary distribution is of course better than $\MCany$, which is evident in the second range of noise.
	\item \textbf{Probability of getting sorted sequence.} We evaluate how the probability of hitting the sorted sequence changes when the noise increases. In \Cref{app:exp},  Figure~\ref{fig:stat-prob} deals with the sequence of ten elements $\{1,2,\ldots, 10\}$, while Figure~\ref{fig:stat-prob-one-outlier} is about the one outlier $\{1,1,1,1,1,1,1,1,1,2\}$.
\end{itemize}

\paragraph*{Acknowledgments.} We are grateful to Lucas Boczkowski, Tom\'{a}\v{s} Gaven\u{c}iak, Francesco Pasquale, and Peter Widmayer for useful discussions. Research supported by SNF (project number 200021\_165524).



\renewcommand\bibsection
{\section*{\refname}\small\renewcommand\bibnumfmt[1]{##1.}}		
\bibliographystyle{plainnat}
\bibliography{noisy-sorting}	

\begin{thebibliography}{18}
\providecommand{\natexlab}[1]{#1}
\providecommand{\url}[1]{\texttt{#1}}
\expandafter\ifx\csname urlstyle\endcsname\relax
  \providecommand{\doi}[1]{doi: #1}\else
  \providecommand{\doi}{doi: \begingroup \urlstyle{rm}\Url}\fi

\bibitem[Alonso et~al.(2004)Alonso, Chassaing, Gillet, Janson, Reingold, and
  Schott]{QuicksortPhaseTrans}
Laurent Alonso, Philippe Chassaing, Florent Gillet, Svante Janson, Edward~M
  Reingold, and Ren{\'e} Schott.
\newblock Quicksort with unreliable comparisons: a probabilistic analysis.
\newblock \emph{Combinatorics, Probability and Computing}, 13\penalty0
  (4-5):\penalty0 419--449, 2004.

\bibitem[Benjamini et~al.(2005)Benjamini, Berger, Hoffman, and
  Mossel]{Benetal05}
Itai Benjamini, Noam Berger, Christopher Hoffman, and Elchanan Mossel.
\newblock {Mixing times of the biased card shuffling and the asymmetric
  exclusion process}.
\newblock \emph{Transactions of the American Mathematical Society},
  357\penalty0 (8):\penalty0 3013--3029, 2005.

\bibitem[Bhakta et~al.(2013)Bhakta, Miracle, Randall, and Streib]{Bhaetal13}
Prateek Bhakta, Sarah Miracle, Dana Randall, and Amanda~Pascoe Streib.
\newblock {Mixing times of Markov chains for self-organizing lists and biased
  permutations}.
\newblock In \emph{Proc. of the 24th Annual ACM-SIAM Symposium on Discrete
  Algorithms (SODA)}, pages 1--15, 2013.

\bibitem[Blume(1993)]{blumeGEB93}
Lawrence~E. Blume.
\newblock The statistical mechanics of strategic interaction.
\newblock \emph{Games and Economic Behavior}, 5:\penalty0 387--424, 1993.

\bibitem[Bradley and Terry(1952)]{BraTer52}
Ralph~Allan Bradley and Milton~E. Terry.
\newblock {Rank Analysis of Incomplete Block Designs: I. The Method of Paired
  Comparisons}.
\newblock \emph{Biometrika}, 39\penalty0 (3/4):\penalty0 324--345, 1952.
\newblock ISSN 00063444.
\newblock URL \url{http://www.jstor.org/stable/2334029}.

\bibitem[Braverman and Mossel(2008)]{braverman2008noisy}
Mark Braverman and Elchanan Mossel.
\newblock Noisy sorting without resampling.
\newblock In \emph{Proceedings of the nineteenth annual ACM-SIAM symposium on
  Discrete algorithms}, pages 268--276. Society for Industrial and Applied
  Mathematics, 2008.

\bibitem[Bubley and Dyer(1999)]{BubDye99}
Russ Bubley and Martin Dyer.
\newblock {Faster random generation of linear extensions}.
\newblock \emph{Discrete mathematics}, 201\penalty0 (1):\penalty0 81--88, 1999.

\bibitem[Diaconis and Ram(2000)]{DiaRam00}
Persi Diaconis and Arun Ram.
\newblock {Analysis of systematic scan Metropolis algorithms using
  Iwahori-Hecke algebra techniques}.
\newblock \emph{Michigan Math. J.}, 48\penalty0 (1):\penalty0 157--190, 2000.

\bibitem[Dyer and Greenhill(1998)]{DyerG98}
Martin~E. Dyer and Catherine~S. Greenhill.
\newblock A more rapidly mixing markov chain for graph colorings.
\newblock \emph{Random Struct. Algorithms}, 13\penalty0 (3-4):\penalty0
  285--317, 1998.

\bibitem[Freidlin and Wentzell(1984)]{FreWen84}
M.~Freidlin and A.D. Wentzell.
\newblock \emph{{Random Perturbations of Dynamical Systems}}.
\newblock New York: Springer Verlag, 1984.

\bibitem[Greenberg et~al.(2009)Greenberg, Pascoe, and Randall]{GreenbergPR09}
Sam Greenberg, Amanda Pascoe, and Dana Randall.
\newblock Sampling biased lattice configurations using exponential metrics.
\newblock In \emph{Proceedings of the Twentieth Annual {ACM-SIAM} Symposium on
  Discrete Algorithms, {SODA} 2009, New York, NY, USA, January 4-6, 2009},
  pages 76--85, 2009.

\bibitem[Hadjicostas and Lakshmanan(2011)]{recursiveMergeSort}
Petros Hadjicostas and KB~Lakshmanan.
\newblock Recursive merge sort with erroneous comparisons.
\newblock \emph{Discrete Applied Mathematics}, 159\penalty0 (14):\penalty0
  1398--1417, 2011.

\bibitem[Kelly(2011)]{Kel11}
Frank~P. Kelly.
\newblock \emph{Reversibility and stochastic networks}.
\newblock Cambridge University Press, 2011.

\bibitem[Levin et~al.(2009)Levin, Peres, and Wilmer]{LevPerWil09}
David~Asher Levin, Yuval Peres, and Elizabeth~Lee Wilmer.
\newblock \emph{{M}arkov chains and mixing times}.
\newblock American Mathematical Soc., 2009.

\bibitem[Mallows(1957)]{Mal57}
Colin~L. Mallows.
\newblock {Non-null ranking models. I}.
\newblock \emph{Biometrika}, 44\penalty0 (1/2):\penalty0 114--130, 1957.

\bibitem[Mezard and Montanari(2009)]{mezard2009information}
Marc Mezard and Andrea Montanari.
\newblock \emph{Information, physics, and computation}.
\newblock Oxford University Press, 2009.

\bibitem[Palem and Lingamneni(2013)]{PalemA13}
Krishna~V. Palem and Avinash Lingamneni.
\newblock {Ten Years of Building Broken Chips: The Physics and Engineering of
  Inexact Computing}.
\newblock \emph{{ACM} Trans. Embedded Comput. Syst.}, 12\penalty0
  (2s):\penalty0 87, 2013.

\bibitem[Wilson(2004)]{Wil04}
David~Bruce Wilson.
\newblock {Mixing times of lozenge tiling and card shuffling Markov chains}.
\newblock \emph{Annals of Applied Probability}, pages 274--325, 2004.

\end{thebibliography}

\newpage	
\appendix
\section{Additional experiments}\label{app:exp}

\begin{figure}[h!]
	\minipage{0.49\textwidth}
	\includegraphics[width=\linewidth]{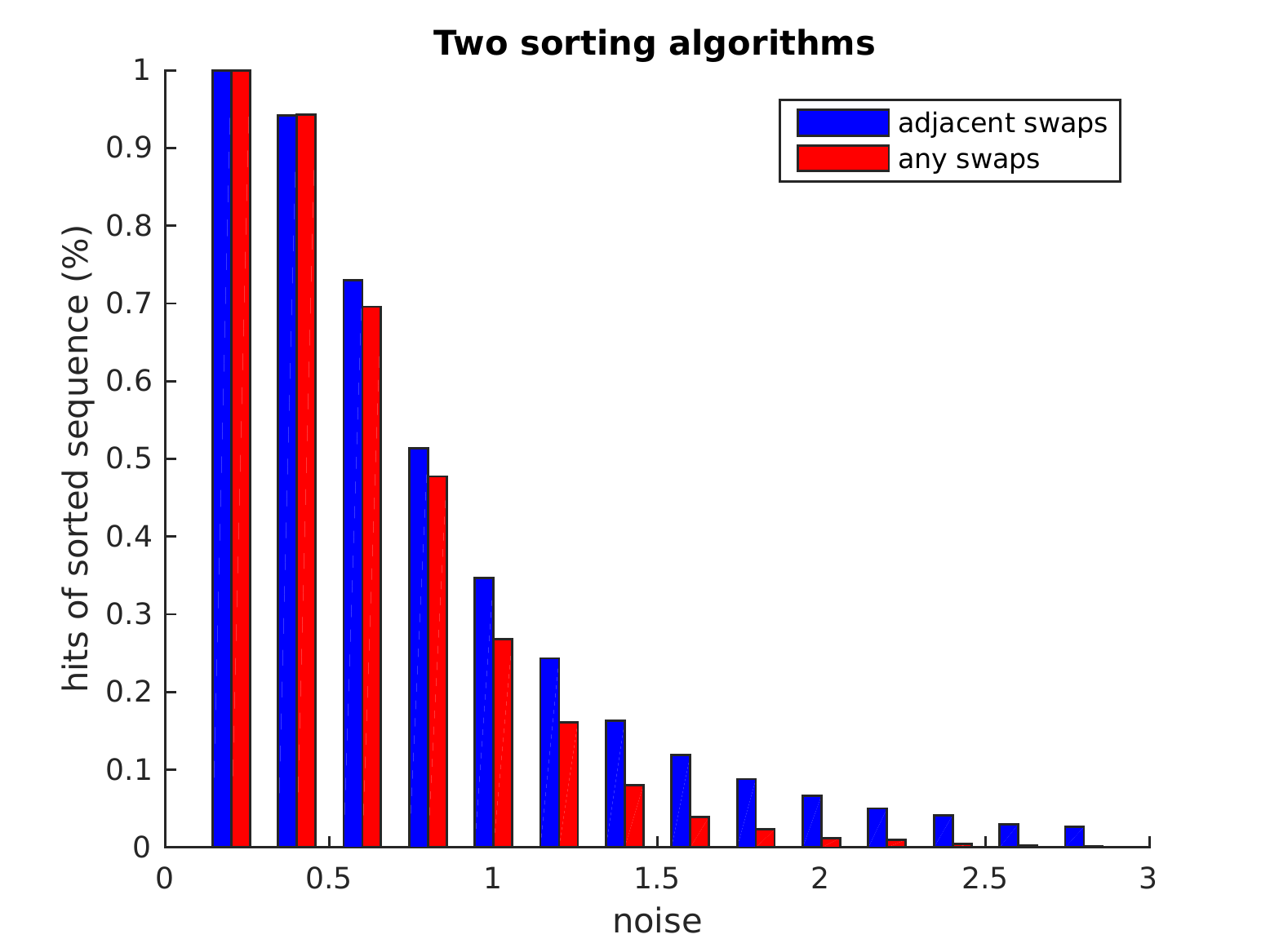}
	\endminipage\hfill
	\minipage{0.47\textwidth}%
	\caption{ 
		We measure the probability of hitting the sorted sequence during a certain number of steps after both algorithms have approached their stationary distribution. The elements to be sorted are $(10,9,\ldots,1)$.  By increasing the value of $noise$, where  $\lambda=e^{1/noise}$,
		the stationary probability of sorted sequence decreases much faster in $\MCany$ than in $\MCadj$.
		}\label{fig:stat-prob}
		\endminipage	
		\end{figure}

		\begin{figure}[h]
			\minipage{0.49\textwidth}
			\includegraphics[width=\linewidth]{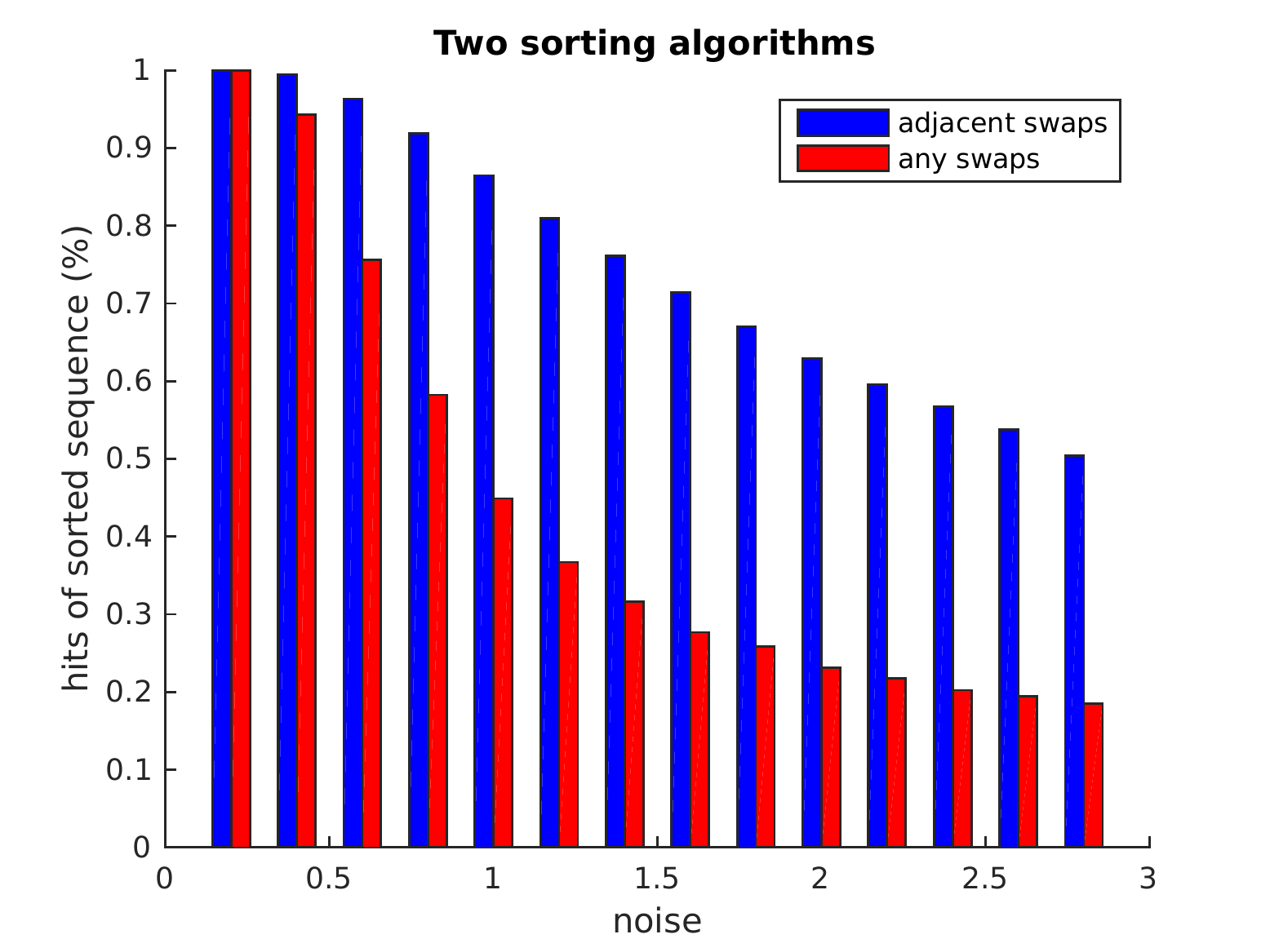}
			\endminipage\hfill
			\minipage{0.49\textwidth}
			\includegraphics[width=\linewidth]{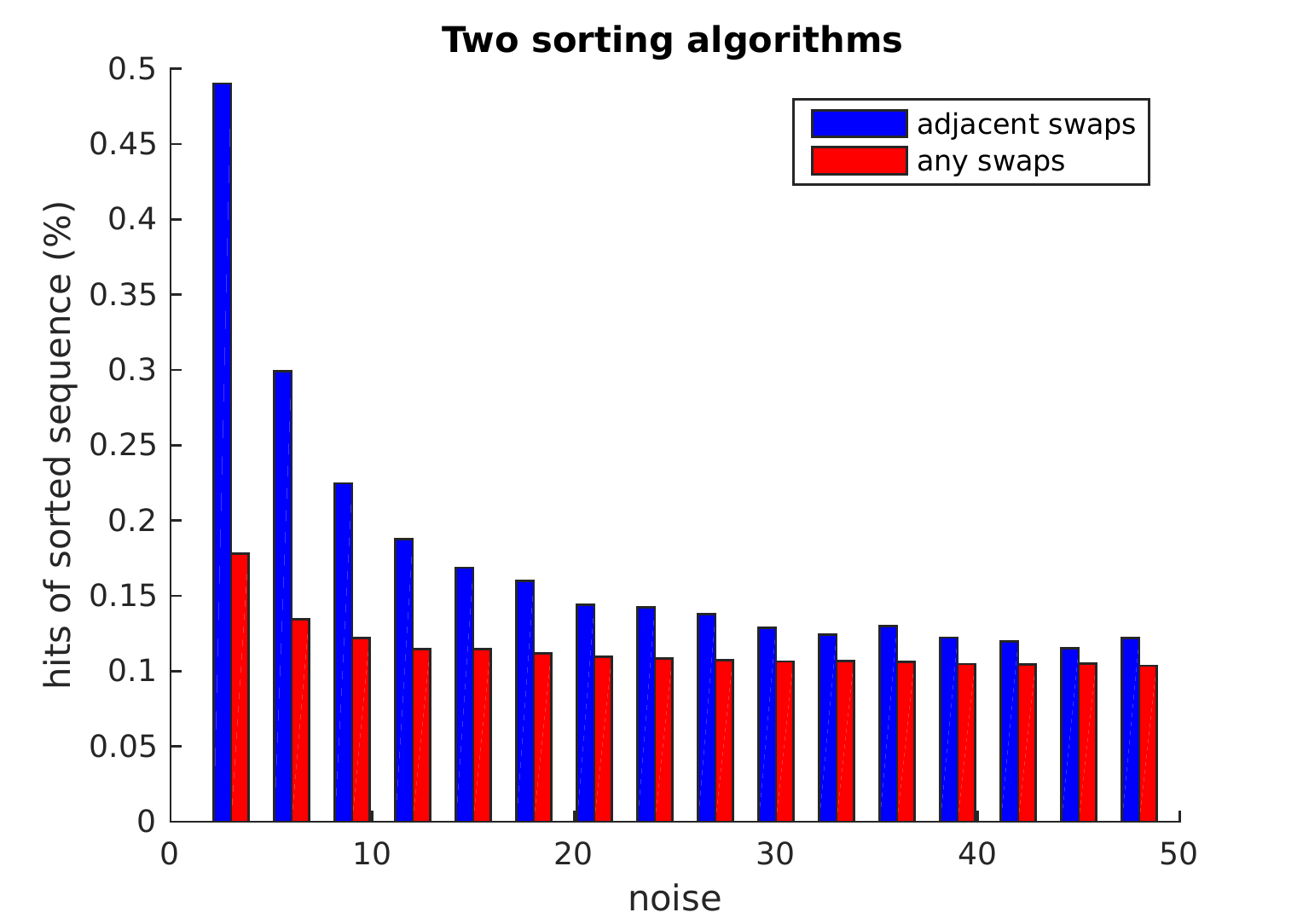}
			\endminipage\hfill
			
			\caption{Percentage of hits of the sorted sequence after both algorithms have reached their stationary distribution. The set of elements is $\{1,1,1,1,1,1,1,1,1,2\}$.}	
			\label{fig:stat-prob-one-outlier}
			\end{figure}

\section{Additional lemmata and proofs}

\subsection{Three elements}\label{app:omitted-proofs}

\setcounter{theorem}{12}
\ratiosabc*
\setcounter{theorem}{16}
\subsubsection{Proof of Lemma~\ref{le:ratios-abc}}\label{app:proof:le:ratios-abc}

We shall use the Markov Chain Tree Theorem \eqref{eq:tree-stationary}. Our goal is thus to show that 
\begin{align}\label{eq:bound:trees-ratios}
	\frac{\sum_{T\in \mathcal T(s) } \P(T)}{\sum_{T'\in \mathcal T(s')} \P(T)} < \frac{\pi_{adj}(s)}{\pi_{adj}(s')} = \lambda^{2(w(s')-w(s))} && \text{for} && w(s) < w(s') \enspace .
\end{align}
The strategy to prove this inequality is to find a suitable mapping from trees $T\in \mathcal{T}(s)$ to trees $T'\in \mathcal{T}(s')$ such that $\frac{\P(T)}{\P(T')} < \frac{\pi_{adj}(s)}{\pi_{adj}(s')}$. The basic mapping consists of reversing the path from $s'$ to $s$ in $T$ to obtain the tree $T'$:

\begin{definition}
	For any $s$ and $s'$, and for any tree $T\in \mathcal{T}(s)$ its \emph{$s'$-reversed} is the tree $T'$ equal to $T$ but with the edges on the path from $s'$ to $s$ reversed.
\end{definition}
Hereafter, we call such a tree $T'$ simply a \textit{reversed} tree of $T$ if $s$ and $s'$ are clear from the context.
We say that an edge has \emph{length} $w$ if its transition probability is of the form $\frac{\lambda^w}{\lambda^w + \lambda^{-w}}$. The \emph{length} of a path is the total length of its edges. The \textit{multiset} of a tree $T$ is the set of absolute edge lengths appearing in $T$,
\[
ms(T) = \{w \mid \text{some edge in $T$ has length $w$ or $-w$}\} \enspace.
\]
Based on this multiset, we denote by $ms(T,w)$ the number of edges in $T$ whose length is either $w$ or $-w$.
\begin{fact}
	Let $T$ be a tree containing a path from $s'$ to $s$ of length $\ell$, and let $T'$ be its reversed tree. Then the probabilities of these two trees are of the form
	\begin{align}\label{eq:goodtree}
		\P(T) =& \frac{\lambda^\ell \cdot \lambda^{L(T)}}{M(T)} & \text{and}& & \P(T') =& \frac{\lambda^{-\ell} \cdot \lambda^{L(T)}}{M(T)}
	\end{align}
	where $L(T)$ denotes the total length of the edges which are in the tree but not in the path between $s$ and $s'$, and $M(T)$ depends on the multiset $ms(T)$, i.e., 
	$$
	M(T) = \prod_{w\in ms(T)} (\lambda^w+ \lambda^{-w})^{ms(T,w)} \enspace .
	$$
\end{fact}	
From \eqref{eq:goodtree} we get $\frac{\P(T)}{\P(T')} \leq \lambda^{2\ell}$. This motivates the following definition.

\begin{definition}
	For any $s$ and $s'$, we call a tree $T\in \mathcal{T}(s)$ \emph{good} if the length $\ell$ of its path from $s'$ to $s$ satisfies $\lambda^{2\ell} \leq \frac{\pi_{adj}(s)}{\pi_{adj}(s')}$. Otherwise we call $T$ a \emph{bad} tree.
\end{definition}

By this definition the basic mapping of a bad tree does not give the desired bound \eqref{eq:bound:trees-ratios}. In such cases, we will map groups of bad trees into groups of good ones, depending on $s$ and $s'$.

\newcommand{\xabcbac}{{\Edge[label=$-x$](abc)(bac);}}
\newcommand{\xbacabc}{{\Edge[label=$x$](bac)(abc);}}
\newcommand{\xabcacb}{{\Edge[label=$-y$](abc)(acb);}}
\newcommand{\xacbabc}{{\Edge[label=$y$](acb)(abc);}}
\newcommand{\xbacbca}{{\Edge[label=$-x-y$](bac)(bca);}}
\newcommand{\xbcabac}{{\Edge[label=$x+y$](bca)(bac);}}
\newcommand{\xbcacba}{{\Edge[label=$-y$](bca)(cba);}}
\newcommand{\xcbabca}{{\Edge[label=$y$](cba)(bca);}}
\newcommand{\xcabcba}{{\Edge[label=$-x$](cab)(cba);}}
\newcommand{\xcbacab}{{\Edge[label=$x$](cba)(cab);}}
\newcommand{\xcabacb}{{\Edge[label=$x+y$](cab)(acb);}}
\newcommand{\xacbcab}{{\Edge[label=$-x-y$](acb)(cab);}}
\newcommand{\xacbbca}{{\Edge[label=$-x$, style={bend left}
		](acb)(bca);}}
\newcommand{\xbcaacb}{{\Edge[label=$x$, style={bend left}
		](bca)(acb);}}
\newcommand{\xabccba}{{\Edge[label=$-x-y$](abc)(cba);}}
\newcommand{\xcbaabc}{{\Edge[label=$x+y$](cba)(abc);}}
\newcommand{\xbaccab}{{\Edge[label=$-y$, style={bend left}
		](bac)(cab);}}
\newcommand{\xcabbac}{{\Edge[label=$y$, style={bend left}
		](cab)(bac);}}

\paragraph{The (bac) vs  (bca) case. } Observe that $\frac{\pi_{adj}(bac)}{\pi_{adj}(bca)}= \lambda^{2(c-a)}$. In this case there is no bad tree, since all paths from (bca) to (bac) have length at most $c-a$ (see Figure~\ref{fig:bcabac}). 

\begin{figure}[h]
	\begin{subfigure}[b]{0.32\textwidth}
		\centering
		\resizebox{\linewidth}{!}{
			\begin{tikzpicture}
			\states
			
			\bcabac
			\end{tikzpicture}
		}
		\caption{$\lambda^{c-a}$}
		\label{fig:bcabac:1}
	\end{subfigure}
	\begin{subfigure}[b]{0.32\textwidth}
		\centering
		\resizebox{\linewidth}{!}{
			\begin{tikzpicture}
			\states
			
			\bcaacb \acbabc \abcbac
			\end{tikzpicture}
		}
		\caption{$\lambda^{c-b}$}   
		\label{fig:bcabac:2}
	\end{subfigure}
	\begin{subfigure}[b]{0.32\textwidth}
		\centering
		\resizebox{\linewidth}{!}{
			\begin{tikzpicture}
			\states
			
			\bcaacb \acbabc \abccba \cbacab \cabbac
			\end{tikzpicture}
		}
		\caption{$\lambda^{c-a}$}
		\label{fig:bcabac:3}
	\end{subfigure}
	
	\begin{subfigure}[b]{0.32\textwidth}
		\centering
		\resizebox{\linewidth}{!}{
			\begin{tikzpicture}
			\states
			
			\bcaacb \acbcab \cabbac
			\end{tikzpicture}
		}
		\caption{$\lambda^{0}$}
		\label{fig:bcabac:4}
	\end{subfigure}
	\begin{subfigure}[b]{0.32\textwidth}
		\centering
		\resizebox{\linewidth}{!}{
			\begin{tikzpicture}
			\states
			
			\bcaacb \acbcab \cabcba \cbaabc \abcbac
			\end{tikzpicture}
		}
		\caption{$\lambda^{a-b}$}
		\label{fig:bcabac:5}
	\end{subfigure}
	\begin{subfigure}[b]{0.32\textwidth}
		\centering
		\resizebox{\linewidth}{!}{
			\begin{tikzpicture}
			\states
			
			\bcacba \cbaabc \abcbac
			\end{tikzpicture}
		}
		\caption{$\lambda^{0}$}
		\label{fig:bcabac:6}
	\end{subfigure}
	\begin{subfigure}[b]{0.32\textwidth}
		\centering
		\resizebox{\linewidth}{!}{
			\begin{tikzpicture}
			\states
			
			\bcacba \cbaabc \abcacb \acbcab \cabbac
			\end{tikzpicture}
		}
		\caption{$\lambda^{b-c}$}
		\label{fig:bcabac:7}
	\end{subfigure}
	\begin{subfigure}[b]{0.32\textwidth}
		\centering
		\resizebox{\linewidth}{!}{
			\begin{tikzpicture}
			\states
			
			\bcacba \cbacab \cabbac
			\end{tikzpicture}
		}
		\caption{$\lambda^{b-a}$}
		\label{fig:bcabac:8}
	\end{subfigure}
	\begin{subfigure}[b]{0.32\textwidth}
		\centering
		\resizebox{\linewidth}{!}{
			\begin{tikzpicture}
			\states
			
			\bcacba \cbacab \cabacb \acbabc \abcbac
			\end{tikzpicture}
		}
		\caption{$\lambda^{c-a}$}
		\label{fig:bcabac:9}
	\end{subfigure}
	
	\caption{All paths from (bca) to (bac). There are no bad trees for these two states.} 
	\label{fig:bcabac}
\end{figure}

\paragraph{The (abc) vs (bac) case. }
Note that $\frac{\pi_{adj}(bac)}{\pi_{adj}(bca)}= \lambda^{2(b-a)}$. Figure~\ref{fig:bacabc} shows all paths from $s'=(bac)$ to $s=(abc)$.
Trees using the path in \ref{fig:bacabc:4} are bad, since this path has length $\ell=c-a>b-a$.
Trees using the path in \ref{fig:bacabc:3} are bad if $c-b > b-a$.
We combine the bad trees with the good trees having paths in \ref{fig:bacabc:2}, \ref{fig:bacabc:6}, \ref{fig:bacabc:8}, or \ref{fig:bacabc:9}.

	\begin{figure}
		\begin{subfigure}[b]{0.32\textwidth}
			\centering
			\resizebox{\linewidth}{!}{
				\begin{tikzpicture}
				\states
				
				\bacabc
				\end{tikzpicture}
			}
			\caption{$\lambda^{b-a}$}
			\label{fig:bacabc:1}
		\end{subfigure}
		\begin{subfigure}[b]{0.32\textwidth}
			\centering
			\resizebox{\linewidth}{!}{
				\begin{tikzpicture}
				\states
				
				\bacbca \bcaacb \acbabc
				\end{tikzpicture}
			}
			\caption{$\lambda^{0}$}   
			\label{fig:bacabc:2}
		\end{subfigure}
		\begin{subfigure}[b]{0.32\textwidth}
			\centering
			\resizebox{\linewidth}{!}{
				\begin{tikzpicture}
				\states
				
				\baccab \cabcba \cbabca \bcaacb \acbabc
				\end{tikzpicture}
			}
			\caption{$\lambda^{c-b}$}
			\label{fig:bacabc:3}
		\end{subfigure}
		
		\begin{subfigure}[b]{0.32\textwidth}
			\centering
			\resizebox{\linewidth}{!}{
				\begin{tikzpicture}
				\states
				
				\baccab \cabacb \acbabc
				\end{tikzpicture}
			}
			\caption{$\lambda^{c-a}$}
			\label{fig:bacabc:4}
		\end{subfigure}
		\begin{subfigure}[b]{0.32\textwidth}
			\centering
			\resizebox{\linewidth}{!}{
				\begin{tikzpicture}
				\states
				
				\bacbca \bcacba \cbacab \cabacb \acbabc
				\end{tikzpicture}
			}
			\caption{$\lambda^{b-a}$}
			\label{fig:bacabc:5}
		\end{subfigure}
		\begin{subfigure}[b]{0.32\textwidth}
			\centering
			\resizebox{\linewidth}{!}{
				\begin{tikzpicture}
				\states
				
				\bacbca \bcacba \cbaabc
				\end{tikzpicture}
			}
			\caption{$\lambda^{b-c}$}
			\label{fig:bacabc:6}
		\end{subfigure}
		\begin{subfigure}[b]{0.32\textwidth}
			\centering
			\resizebox{\linewidth}{!}{
				\begin{tikzpicture}
				\states
				
				\baccab \cabacb \acbbca \bcacba \cbaabc
				\end{tikzpicture}
			}
			\caption{$\lambda^{b-a}$}
			\label{fig:bacabc:7}
		\end{subfigure}
		\begin{subfigure}[b]{0.32\textwidth}
			\centering
			\resizebox{\linewidth}{!}{
				\begin{tikzpicture}
				\states
				
				\baccab \cabcba \cbaabc
				\end{tikzpicture}
			}
			\caption{$\lambda^{0}$}
			\label{fig:bacabc:8}
		\end{subfigure}
		\begin{subfigure}[b]{0.32\textwidth}
			\centering
			\resizebox{\linewidth}{!}{
				\begin{tikzpicture}
				\states
				
				\bacbca \bcaacb \acbcab \cabcba \cbaabc
				\end{tikzpicture}
			}
			\caption{$\lambda^{a-c}$}
			\label{fig:bacabc:9}
		\end{subfigure}
		
		\caption{All paths from (bac) to (abc). Bad trees include path \subref{fig:bacabc:3} or \subref{fig:bacabc:4}.} 
		\label{fig:bacabc}
	\end{figure}

\begin{lemma}\label{lem:badgood}
	Let $\mathcal{BAD}(s)\subset \mathcal{T}(s)$ be the set of bad trees with $s's$-path \ref{fig:bacabc:3} or \ref{fig:bacabc:4}, and let $\mathcal{GOOD}(s)\subset \mathcal{T}(s)$ be the set of good trees with \ref{fig:bacabc:2}, \ref{fig:bacabc:6}, \ref{fig:bacabc:8} or  \ref{fig:bacabc:9}. Then,		
	\begin{align}\label{eq:allgoodandbadtrees}
		\frac{\sum_{T\in \mathcal{BAD}(s)}\P(T) + \sum_{T\in\mathcal{GOOD}(s)}\P(T)}{\sum_{T\in \mathcal{BAD}(s)}\P( T') + \sum_{T\in \mathcal{GOOD}(s)}\P( T')} \leq \lambda^{2(b-a)} \enspace .
	\end{align}
\end{lemma}
\begin{proof}
	Let us use $x:= b-a$, $y:= c-b$, and $x+y := c-a$.
	The trees and their probabilities are shown in Figures \ref{fig:bad3good9trees}-\ref{fig:goodtreesH} (in Appendix \ref{app:figs}).
	We will show {\small
		$$ \left(\sum_{T\in \mathcal{BAD}(s)}\P(T) + \sum_{T\in\mathcal{GOOD}(s)}\P(T) \right) \leq \lambda^{2x} \left(\sum_{T\in \mathcal{BAD}(s)}\P( T') + \sum_{T\in \mathcal{GOOD}(s)}\P( T') \right)$$}%
	which obviously implies (\ref{eq:allgoodandbadtrees}).
	To get rid of the fractions in the probabilities we multiply them by the least common multiple of their denominators, i.e., $$\text{LCM} := (\lambda^{x+y}+\lambda^{-x-y})^3(\lambda^{y}+\lambda^{-y})^3(\lambda^{x}+\lambda^{-x})^2 \enspace .$$ We get for the left hand side of the inequality
	{ 
		\begin{align*}
			\text{LCM}   \left(\sum_{T\in \mathcal{BAD}(s)}\P( T) + \sum_{T\in \mathcal{GOOD}(s)}\P( T) \right)  = &\enspace			
			6 {\lambda^{5 x+4 y}}
			+16 {\lambda^{3 x+4 y}}
			+10 {\lambda^{x+4 y}}
			+6 {\lambda^{5 x+2 y}}\\
			&+26 {\lambda^{3 x+2 y}}
			+42 {\lambda^{x+2 y}}
			+18 {\lambda^{-x+2 y}}
			+8 {\lambda^{x-2 y}}\\
			&+24 {\lambda^{-x-2 y}}
			+12 {\lambda^{-3 x-2 y}}
			+4 {\lambda^{-x-4 y}}
			+4 {\lambda^{-3 x-4 y}}\\
			&+10 {\lambda^{3 x}}
			+36 {\lambda^{x}}
			+42 {\lambda^{-x}}
			+8{\lambda^{-3 x}}\enspace .	
		\end{align*}
	}
	And we get for the right hand side
	{ 
		\begin{align*}
			\lambda^{2x}  \text{LCM}    \left(\sum_{T\in \mathcal{BAD}(s)}\P( T') + \sum_{T\in \mathcal{GOOD}(s)}\P( T')\right) = &\enspace
			10 {\lambda^{5 x+4 y}}
			+16 {\lambda^{3 x+4 y}}
			+6 {\lambda^{x+4 y}}
			+18 {\lambda^{5 x+2 y}}\\
			&+42 {\lambda^{3 x+2 y}}
			+26 {\lambda^{x+2 y}}
			+6 {\lambda^{-x+2 y}}
			+8 {\lambda^{5 x}}\\
			&+42 {\lambda^{3 x}}
			+36 {\lambda^{x}}
			+10 {\lambda^{-x}}		
			+12 {\lambda^{3 x-2 y}}\\
			&+24 {\lambda^{x-2 y}}
			+8 {\lambda^{-x-2 y}}
			+4 {\lambda^{x-4 y}}
			+4 {\lambda^{-x-4 y}} \enspace .
		\end{align*}
	}
	The difference between the left and the right hand side is
	{ 
		\begin{align*}
			4 {\lambda^{5 x+4 y}}
			-4 {\lambda^{x+4 y}}
			+12 {\lambda^{5 x+2 y}}
			-12 {\lambda^{-x+2 y}}
			+16 {\lambda^{3 x+2 y}}
			-16 {\lambda^{x+2 y}}
			+8 {\lambda^{5 x}}
			-8{\lambda^{-3 x}}\\
			+32 {\lambda^{3 x}}
			-32{\lambda^{-x}}
			+12 {\lambda^{3 x-2 y}}
			-12 {\lambda^{-3 x-2 y}}
			+16 {\lambda^{x-2 y}}		
			-16 {\lambda^{-x-2 y}}
			+4 {\lambda^{x-4 y}}
			-4 {\lambda^{-3 x-4 y}}
		\end{align*}
	}
	and we can pair up the terms to conclude that this difference is positive. 
\end{proof}

\paragraph{The (bca) vs (cba) case. }	
Note that $\frac{\pi_{adj}(bca)}{\pi_{adj}(cba)}= \lambda^{2(c-b)}$. 
Figure~\ref{fig:bcacba} in Appendix \ref{app:figs} shows all paths from $s'=(cba)$ to $s=(bca)$.
We can combine the bad trees with $s's$-path \ref{fig:bcacba:4} or \ref{fig:bcacba:5}, and the good trees with paths \ref{fig:bcacba:3}, \ref{fig:bcacba:6}, \ref{fig:bcacba:8} or \ref{fig:bcacba:9} to show that the ratio between all trees and their reversals is smaller than $\lambda^{2(c-b)}$:
\begin{lemma}\label{lem:similarlemma}
	For the bad trees $\mathcal{BAD}(s)\subset \mathcal{T}(a)$ with $s's$-path \ref{fig:bcacba:4} or \ref{fig:bcacba:5}, and the good trees $\mathcal{GOOD}(s)\subset \mathcal{T}(a)$ with paths \ref{fig:bcacba:3}, \ref{fig:bcacba:6}, \ref{fig:bcacba:8} or \ref{fig:bcacba:9} it holds that
	\begin{align*} 
	\frac{\sum_{T\in \mathcal{BAD}(s)}\P( T) + \sum_{T\in \mathcal{GOOD}(s)}\P( T)}{\sum_{T\in \mathcal{BAD}(s)}\P( T') + \sum_{T\in \mathcal{GOOD}(s)}\P( T')} \leq \lambda^{2(c-b)} \enspace .
	\end{align*}
\end{lemma}
\begin{proof}
	We proceed as in the proof for Lemma \ref{lem:badgood} and finally get the difference between the left and the right hand side:
	{ 
		\begin{align*}
		4 {\lambda^{4 x +5 y}}
		-4 {\lambda^{4 x + y}}
		+12 {\lambda^{2 x+5 y}}
		-12 {\lambda^{2 x- y}}
		+16 {\lambda^{2 x + 3 y}}		
		-16 {\lambda^{2 x+ y}}
		+8 {\lambda^{5 y}}
		-8{\lambda^{-3 y}}\\
		+32 {\lambda^{3 y}}
		-32{\lambda^{-y}}
		+12 {\lambda^{-2 x+3 y}}
		-12 {\lambda^{-2 x-3 y}}
		+16 {\lambda^{-2 x+ y}}
		-16 {\lambda^{-2 x+- y}}
		+ 4 {\lambda^{-4 x +y}}
		-4 {\lambda^{-4 x -3 y}}
		\enspace 
		\end{align*}
	}
\end{proof}

\paragraph{The other cases. }
There are three other pairs of states for which we need to prove Lemma \ref{le:ratios-abc}. However, the procedure is always the same: We identify the bad and good trees and combine them to conclude \eqref{eq:lemma1}. For the missing paths consider Figures \ref{fig:acbabc}-\ref{fig:cbacab} in Appendix \ref{app:figs}.

\subsection{One outlier}\label{app:simple-chain}

\setcounter{lemma}{14}
\statdist

\begin{proof}
	We get the stationary distribution for $\MCadj$ using the global balance condition of stationary distribution\footnote{The stationary distribution $\pi$ of a Markov chain with transition matrix $P$ must satisfy $\sum_{s'\neq s}\pi(s)P(s,s') = \sum_{s'\neq s} \pi(s')P(s',s)$, for any two states $s,s'$.}, which implies that $\pi_{adj}(s^{(i)}) = \pi_{adj}(s^{(i+1)})(\frac{1-p}{p})$. By the structure of our Markov chain we get that for any state $s^{(j)}$ it holds that 
	$$\pi_{adj}(s^{(j)}) = \pi_{adj}(s^{(i)})(\frac{1-p}{p})^{i-j}.$$ 
	Since the sum of all $\pi_{adj}$ is one, we can express $\pi_{adj}(s^{(i)})$ as
	\begin{align*}
	\pi_{adj}(s^{(i)}) &=  1 - \pi_{adj}(s^{(i)})\left(\sum_{j=1}^{i-1} (\frac{p}{1-p})^j + \sum_{j=1}^{n-i}(\frac{1-p}{p})^{j}\right)\\
	&= \frac{1}{1+\sum_{j=1}^{i-1} (\frac{p}{1-p})^j + \sum_{j=1}^{n-i}(\frac{1-p}{p})^{j}}.	
	\intertext{By applying the formula for geometric series, we rewrite}
	\frac{1}{\pi_{adj}(s^{(i)})} 
	&= 1+ \frac{1-(\frac{p}{1-p})^{i}}{1-(\frac{p}{1-p})}-1 + \frac{1-(\frac{1-p}{p})^{n-i+1}}{1-(\frac{1-p}{p})}-1\\
	&= -1 + \frac{(1-p)^i-p^i}{(1-2p)(1-p)^{i-1}} + \frac{p^{n-i+1}-(1-p)^{n-i+1}}{(2p-1)p^{n-i}}
	\intertext{We expand all terms to the same denominator:}
	&=\frac{-(1-2p)(1-p)^{i-1} p^{n-i} +  (1-p)^{i}p^{n-i} - p^n - (1-p)^{i-1}p^{n-i+1} + (1-p)^{n} }{(1-2p)(1-p)^{i-1}p^{n-i}}
	\intertext{Finally, we can use that $(1-2p)=(1-p)-p$ and observe that the fraction simplifies to what we claimed.}
	\end{align*}
		
	The formula for the stationary distribution of $\MCany$ can be derived as follows using the global balance condition of stationary distribution. For any $\pi_{any}(s^{(i)})$ it holds that
	\begin{align*}
	\pi_{any}(s^{(i)}) \left((i-1)p + (n-i)(1-p)\right) &= (1-p) \sum_{j=1}^{i-1}\pi_{any}(s^{(j)}) + p \sum_{j=i+1}^n \pi_{any}(s^{(j)})\\
	&= (1-p) \sum_{j=1}^{i-1}\pi_{any}(s^{(j)}) + p \left(1- \sum_{j=1}^{i}\pi_{any}(s^{(j)})\right)
	\intertext{that is}
	\pi_{any}(s^{(i)}) \left(ip + (n-i)(1-p)\right) &= (1-2p) \sum_{j=1}^{i-1}\pi_{any}(s^{(j)}) + p\enspace.
	\end{align*}
	Now we can show by induction on $i$ that this recurrence resolves to $$\pi_{any}(s^{(i)}) = \frac{np(1-p)}{((n-i+1)(1-p)+(i-1)p)((n-i)(1-p)+ip)}.$$
	For $i=1$ we immediately get
	$\pi_{any}(s^{(1)}) \left(p + (n-1)(1-p) \right) = p $, which we rewrite as
	$$\pi_{any}(s^{(1)}) = \frac{p}{(n-1)(1-p)+p} = \frac{np(1-p)}{(n(1-p))((n-1)(1-p)+p)}.$$
	For $i>1$ we assume that the formula holds for $i-1$ and we get
	\begin{align*}
	\pi_{any}(s^{(i)}) &= \frac{(1-2p) \sum_{j=1}^{i-1}\pi_{any}(s^{(j)}) + p}{ \left(ip + (n-i)(1-p)\right)}\\
	&= \frac{(1-2p) \sum_{j=1}^{i-2}\pi_{any}(s^{(j)}) + p + \pi_{any}(s^{(i-1)})(1-2p)}{ \left(ip + (n-i)(1-p)\right)}\\
	&= \frac{\pi_{any}(s^{(i-1)})((i-1)p + (n-i+1)(1-p)) + \pi_{any}(s^{(i-1)})(1-2p)}{ \left(ip + (n-i)(1-p)\right)}\\
	&= \frac{\pi_{any}(s^{(i-1)})((i-2)p + (n-i+2)(1-p))}{\left(ip + (n-i)(1-p)\right)}\\
	&= \frac{np(1-p)}{((n-i+1)(1-p)+(i-1)p)\left(ip + (n-i)(1-p)\right)}. \hspace{2cm} 
	\end{align*}	
\end{proof}

\expweight
\setcounter{lemma}{20}
\begin{proof}
We apply the generic formula for the expected weighted inversion \eqref{eq:expected_diclos:generic} to derive the expected weighted inversion of $\MCadj$ and $\MCany$. Since $\pi_{adj}(s^{(i)}) =\pi_{adj}(s^{(n)}) \left(\frac{p}{1-p}\right)^{n-i}$ we get
\begin{align*}
E^{w}_{adj} &= \sum_{i=1}^n (n-i)(x-1)\pi_{adj}(s^{(i)})\\
&= (x-1)\pi_{adj}(s^{(n)})\cdot\sum_{i=0}^{n-1}i \left(\frac{p}{1-p}\right)^i\\
&= (x-1) \frac{(1-p)^{n-1}(1-2p)}{(1-p)^n-p^n}\cdot\frac{(n-1)\left(\frac{p}{1-p}\right)^{n-1}-n\left(\frac{p}{1-p}\right)^n+\left(\frac{p}{1-p}\right)}{\left(\frac{p}{1-p}-1\right)^2}\\
&= n(x-1)p\left(\frac{1}{n(1-2p)}-\frac{p^{n-1}}{(1-p)^n-p^n}\right)\enspace.
\end{align*}
Then observe that for $0<p<\frac{1}{2}$, the first inequality is immediate from $\frac{p^n}{(1-p)^n - p^n}>0$.
As for $\MCany$ we have
\begin{align*}
E^{w}_{any} &= \sum_{i=1}^n (n-i)(x-1)\pi_{any}(s^{(i)})\\
&= (x-1)\cdot \sum_{i=0}^{n-1}i\cdot\pi_{any}(s^{(n-i)})\\
&= n(x-1)p\cdot \sum_{i=0}^{n-1} \frac{i(1-p)}{((i+1)(1-p)+(n-i-1)p)(i(1-p)+(n-i)p)}\enspace.
\end{align*}
Observe that we can lower bound the sum in the formula by the integral
$$\int_{0}^{n}\frac{i(1-p)}{((i+1)(1-p)+(n-i-1)p)(i(1-p)+(n-i)p)}di\enspace,$$
which is larger than 1 if $0<p<\frac{1}{2}$. 
\end{proof}

~

~

~

\section{Additional pictures}\label{app:figs}

		\begin{figure}[h!]
			\centering
			\begin{minipage}[b]{0.49\textwidth}
				\centering
				\resizebox{.8\linewidth}{!}{
					\begin{tikzpicture}
					\states
					
					\xbaccab \xcabcba \xcbabca \xbcaacb \xacbabc
					\end{tikzpicture}
				}
				\label{fig:bad9}
				\subcaption*{$\P(T) = \frac{\lambda^{y}}{(\lambda^{x}+\lambda^{-x})^2(\lambda^{y}+\lambda^{-y})^3}$\\
					$\P(T') = \frac{\lambda^{-y}}{(\lambda^{x}+\lambda^{-x})^2(\lambda^{y}+\lambda^{-y})^3}$\\}
			\end{minipage}	
			\begin{minipage}[b]{0.49\textwidth}
				\centering
				\resizebox{.8\linewidth}{!}{
					\begin{tikzpicture}
					\states
					
					\xbacbca \xbcaacb \xacbcab \xcabcba \xcbaabc
					\end{tikzpicture}
				}
				\label{fig:good25}
				\subcaption*{$\P(T) = \frac{\lambda^{-x-y}}{(\lambda^{x+y}+\lambda^{-x-y})^3(\lambda^{x}+\lambda^{-x})^2}$\\
					$\P(T') = \frac{\lambda^{x+t}}{(\lambda^{x+y}+\lambda^{-x-y})^3(\lambda^{x}+\lambda^{-x})^2}$\\}
			\end{minipage}
			
			\caption{Bad tree with path \ref{fig:bacabc:3} and good tree with path \ref{fig:bacabc:9}. }
			\label{fig:bad3good9trees}
		\end{figure}
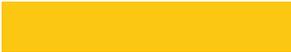
	
	\begin{figure}[h]
		\centering
		\begin{minipage}[b]{0.49\textwidth}
			\centering
			\resizebox{.8\linewidth}{!}{
				\begin{tikzpicture}
				\states
				
				\xbaccab \xcabacb \xacbabc \xcbabca \xbcabac
				\end{tikzpicture}
			}
			\label{fig:bad1}
			\subcaption*{$\P(T) = \frac{\lambda^{2x+3y}}{(\lambda^{x+y}+\lambda^{-x-y})^2(\lambda^{y}+\lambda^{-y})^3}$\\
				$\P(T') = \frac{\lambda^{y}}{(\lambda^{x+y}+\lambda^{-x-y})^2(\lambda^{y}+\lambda^{-y})^3}$\\}
		\end{minipage}	
		\begin{minipage}[b]{0.49\textwidth}
			\centering
			\resizebox{.8\linewidth}{!}{
				\begin{tikzpicture}
				\states
				
				\xbaccab \xcabacb \xacbabc \xcbacab \xbcabac
				\end{tikzpicture}
			}
			\label{fig:bad2}
			\subcaption*{$\P(T) = \frac{\lambda^{3x+2y}}{(\lambda^{x+y}+\lambda^{-x-y})^2(\lambda^{y}+\lambda^{-y})^2(\lambda^{x}+\lambda^{-x})}$\\
				$\P(T') = \frac{\lambda^{x}}{(\lambda^{x+y}+\lambda^{-x-y})^2(\lambda^{y}+\lambda^{-y})^2(\lambda^{x}+\lambda^{-x})}$\\}
		\end{minipage}
		
		\begin{minipage}[b]{0.49\textwidth}
			\centering
			\resizebox{.8\linewidth}{!}{
				\begin{tikzpicture}
				\states
				
				\xbaccab \xcabacb \xacbabc \xcbacab \xbcacba
				\end{tikzpicture}
			}
			\label{fig:bad3}
			\subcaption*{$\P(T) = \frac{\lambda^{2x}}{(\lambda^{x+y}+\lambda^{-x-y})(\lambda^{y}+\lambda^{-y})^3(\lambda^{x}+\lambda^{-x})}$\\
				$\P(T') = \frac{\lambda^{-2y}}{(\lambda^{x+y}+\lambda^{-x-y})(\lambda^{y}+\lambda^{-y})^3(\lambda^{x}+\lambda^{-x})}$\\}
		\end{minipage}	
		\begin{minipage}[b]{0.49\textwidth}
			\centering
			\resizebox{.8\linewidth}{!}{
				\begin{tikzpicture}
				\states
				
				\xbaccab \xcabacb \xacbabc \xcbaabc \xbcaacb
				\end{tikzpicture}
			}
			\label{fig:bad4}
			\subcaption*{$\P(T) = \frac{\lambda^{3x+2y}}{(\lambda^{x+y}+\lambda^{-x-y})^2(\lambda^{y}+\lambda^{-y})^2(\lambda^{x}+\lambda^{-x})}$\\
				$\P(T') = \frac{\lambda^{x}}{(\lambda^{x+y}+\lambda^{-x-y})^2(\lambda^{y}+\lambda^{-y})^2(\lambda^{x}+\lambda^{-x})}$\\}
		\end{minipage}
		
		\begin{minipage}[b]{0.49\textwidth}
			\centering
			\resizebox{.8\linewidth}{!}{
				\begin{tikzpicture}
				\states
				
				\xbaccab \xcabacb \xacbabc \xcbabca \xbcaacb
				\end{tikzpicture}
			}
			\label{fig:bad5}
			\subcaption*{$\P(T) = \frac{\lambda^{2x+2y}}{(\lambda^{x+y}+\lambda^{-x-y})(\lambda^{y}+\lambda^{-y})^3(\lambda^{x}+\lambda^{-x})}$\\
				$\P(T') = \frac{\lambda^{0}}{(\lambda^{x+y}+\lambda^{-x-y})(\lambda^{y}+\lambda^{-y})^3(\lambda^{x}+\lambda^{-x})}$\\}
		\end{minipage}
		\begin{minipage}[b]{0.49\textwidth}
			\centering
			\resizebox{.8\linewidth}{!}{
				\begin{tikzpicture}
				\states
				
				\xbaccab \xcabacb \xacbabc \xcbacab \xbcaacb
				\end{tikzpicture}
			}
			\label{fig:bad6}
			\subcaption*{$\P(T) = \frac{\lambda^{3x+y}}{(\lambda^{x+y}+\lambda^{-x-y})(\lambda^{y}+\lambda^{-y})^2(\lambda^{x}+\lambda^{-x})^2}$\\
				$\P(T') = \frac{\lambda^{x-y}}{(\lambda^{x+y}+\lambda^{-x-y})(\lambda^{y}+\lambda^{-y})^2(\lambda^{x}+\lambda^{-x})^2}$\\}
		\end{minipage}
		
		\begin{minipage}[b]{0.49\textwidth}
			\centering
			\resizebox{.8\linewidth}{!}{
				\begin{tikzpicture}
				\states
				
				\xbaccab \xcabacb \xacbabc \xcbaabc \xbcabac
				\end{tikzpicture}
			}
			\label{fig:bad7}
			\subcaption*{$\P(T) = \frac{\lambda^{3x+3y}}{(\lambda^{x+y}+\lambda^{-x-y})^3(\lambda^{y}+\lambda^{-y})^2}$\\
				$\P(T') = \frac{\lambda^{x+y}}{(\lambda^{x+y}+\lambda^{-x-y})^3(\lambda^{y}+\lambda^{-y})^2}$\\}
		\end{minipage}	
		\begin{minipage}[b]{0.49\textwidth}
			\centering
			\resizebox{.8\linewidth}{!}{
				\begin{tikzpicture}
				\states
				
				\xbaccab \xcabacb \xacbabc \xcbaabc \xbcacba
				\end{tikzpicture}
			}
			\label{fig:bad8}
			\subcaption*{$\P(T) = \frac{\lambda^{2x+y}}{(\lambda^{x+y}+\lambda^{-x-y})^2(\lambda^{y}+\lambda^{-y})^3}$\\
				$\P(T') = \frac{\lambda^{-y}}{(\lambda^{x+y}+\lambda^{-x-y})^2(\lambda^{y}+\lambda^{-y})^3}$\\}
		\end{minipage}
		
		\caption{Bad trees with path \ref{fig:bacabc:4}. }
		\label{fig:badtrees}
	\end{figure}
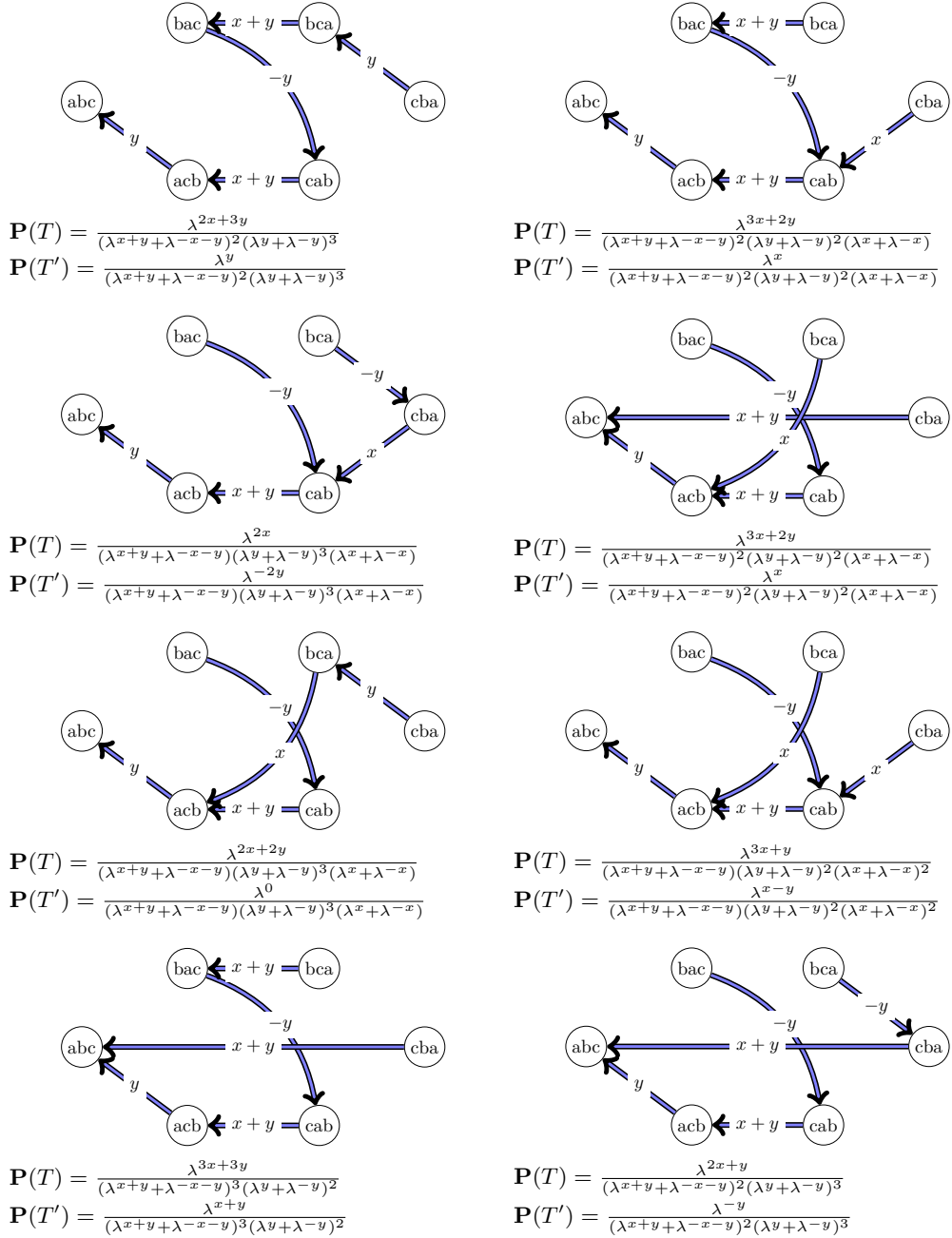

	\begin{figure}[h]
		\centering
		\begin{minipage}[b]{0.49\textwidth}
			\centering
			\resizebox{.8\linewidth}{!}{
				\begin{tikzpicture}
				\states
				
				\xbacbca \xbcaacb \xacbabc \xcbacab \xcabacb
				\end{tikzpicture}
			}  
			\label{fig:good1}
			\subcaption*{$\P(\underline T_1) = \frac{\lambda^{2x+y}}{(\lambda^{x+y}+\lambda^{-x-y})^2(\lambda^{y}+\lambda^{-y})(\lambda^{x}+\lambda^{-x})^2}$\\
				$\P(T') = \frac{\lambda^{2x+y}}{(\lambda^{x+y}+\lambda^{-x-y})^2(\lambda^{y}+\lambda^{-y})(\lambda^{x}+\lambda^{-x})^2}$\\}
		\end{minipage}
		\begin{minipage}[b]{0.49\textwidth}
			\centering
			\resizebox{.8\linewidth}{!}{
				\begin{tikzpicture}
				\states
				
				\xbacbca \xbcaacb \xacbabc \xcbabca \xcabacb
				\end{tikzpicture}
			}   
			\label{fig:good2}
			\subcaption*{$\P(\underline T_2) = \frac{\lambda^{x+2y}}{(\lambda^{x+y}+\lambda^{-x-y})^2(\lambda^{y}+\lambda^{-y})^2(\lambda^{x}+\lambda^{-x})}$\\
				$\P(T') = \frac{\lambda^{x+2y}}{(\lambda^{x+y}+\lambda^{-x-y})^2(\lambda^{y}+\lambda^{-y})^2(\lambda^{x}+\lambda^{-x})}$\\}
		\end{minipage}
		
		\begin{minipage}[b]{0.49\textwidth}
			\centering
			\resizebox{.8\linewidth}{!}{
				\begin{tikzpicture}
				\states
				
				\xbacbca \xbcaacb \xacbabc \xcbabca \xcabcba
				\end{tikzpicture}
			}   
			\label{fig:good3}
			\subcaption*{$\P(\underline T_3) = \frac{\lambda^{-x+y}}{(\lambda^{x+y}+\lambda^{-x-y})(\lambda^{y}+\lambda^{-y})^2(\lambda^{x}+\lambda^{-x})^2}$\\
				$\P(T') = \frac{\lambda^{-x+y}}{(\lambda^{x+y}+\lambda^{-x-y})(\lambda^{y}+\lambda^{-y})^2(\lambda^{x}+\lambda^{-x})^2}$\\}
		\end{minipage}
		\begin{minipage}[b]{0.49\textwidth}
			\centering
			\resizebox{.8\linewidth}{!}{
				\begin{tikzpicture}
				\states
				
				\xbacbca \xbcaacb \xacbabc \xcbaabc \xcabbac
				\end{tikzpicture}
			}   
			\label{fig:good4}
			\subcaption*{$\P(\underline T_4) = \frac{\lambda^{-x+y}}{(\lambda^{x+y}+\lambda^{-x-y})(\lambda^{y}+\lambda^{-y})^2(\lambda^{x}+\lambda^{-x})^2}$\\
				$\P(T') = \frac{\lambda^{-x+y}}{(\lambda^{x+y}+\lambda^{-x-y})(\lambda^{y}+\lambda^{-y})^2(\lambda^{x}+\lambda^{-x})^2}$\\}
		\end{minipage}		
		
		\begin{minipage}[b]{0.49\textwidth}
			\centering
			\resizebox{.8\linewidth}{!}{
				\begin{tikzpicture}
				\states
				
				\xbacbca \xbcaacb \xacbabc \xcbabca \xcabbac
				\end{tikzpicture}
			}   
			\label{fig:good5}
			\subcaption*{$\P(\underline T_5) = \frac{\lambda^{2y}}{(\lambda^{x+y}+\lambda^{-x-y})(\lambda^{y}+\lambda^{-y})^3(\lambda^{x}+\lambda^{-x})}$\\
				$\P(T') = \frac{\lambda^{2y}}{(\lambda^{x+y}+\lambda^{-x-y})(\lambda^{y}+\lambda^{-y})^3(\lambda^{x}+\lambda^{-x})}$\\}
		\end{minipage}
		\begin{minipage}[b]{0.49\textwidth}
			\centering
			\resizebox{.8\linewidth}{!}{
				\begin{tikzpicture}
				\states
				
				\xbacbca \xbcaacb \xacbabc \xcbacab \xcabbac
				\end{tikzpicture}
			}   
			\label{fig:good6}
			\subcaption*{$\P(\underline T_6) = \frac{\lambda^{x+y}}{(\lambda^{x+y}+\lambda^{-x-y})(\lambda^{y}+\lambda^{-y})^2(\lambda^{x}+\lambda^{-x})^2}$\\
				$\P(T') = \frac{\lambda^{x+y}}{(\lambda^{x+y}+\lambda^{-x-y})(\lambda^{y}+\lambda^{-y})^2(\lambda^{x}+\lambda^{-x})^2}$\\}
		\end{minipage}	
		
		\begin{minipage}[b]{0.49\textwidth}
			\centering
			\resizebox{.8\linewidth}{!}{
				\begin{tikzpicture}
				\states
				
				\xbacbca \xbcaacb \xacbabc \xcbaabc \xcabacb
				\end{tikzpicture}
			}   
			\label{fig:good7}
			\subcaption*{$\P(\underline T_7) = \frac{\lambda^{2x+2y}}{(\lambda^{x+y}+\lambda^{-x-y})^3(\lambda^{y}+\lambda^{-y})(\lambda^{x}+\lambda^{-x})}$\\
				$\P(T') = \frac{\lambda^{2x+2y}}{(\lambda^{x+y}+\lambda^{-x-y})^3(\lambda^{y}+\lambda^{-y})(\lambda^{x}+\lambda^{-x})}$\\}
		\end{minipage}
		\begin{minipage}[b]{0.49\textwidth}
			\centering
			\resizebox{.8\linewidth}{!}{
				\begin{tikzpicture}
				\states
				
				\xbacbca \xbcaacb \xacbabc \xcabcba \xcbaabc
				\end{tikzpicture}
			}   
			\label{fig:good8}
			\subcaption*{$\P(\underline T_8) = \frac{\lambda^{y}}{(\lambda^{x+y}+\lambda^{-x-y})^2(\lambda^{y}+\lambda^{-y})(\lambda^{x}+\lambda^{-x})^2}$\\
				$\P(T') = \frac{\lambda^{y}}{(\lambda^{x+y}+\lambda^{-x-y})^2(\lambda^{y}+\lambda^{-y})(\lambda^{x}+\lambda^{-x})^2}$\\}
		\end{minipage}		
		\caption{Good trees with path \ref{fig:bacabc:2}.}
		\label{fig:goodtreesB}
	\end{figure}
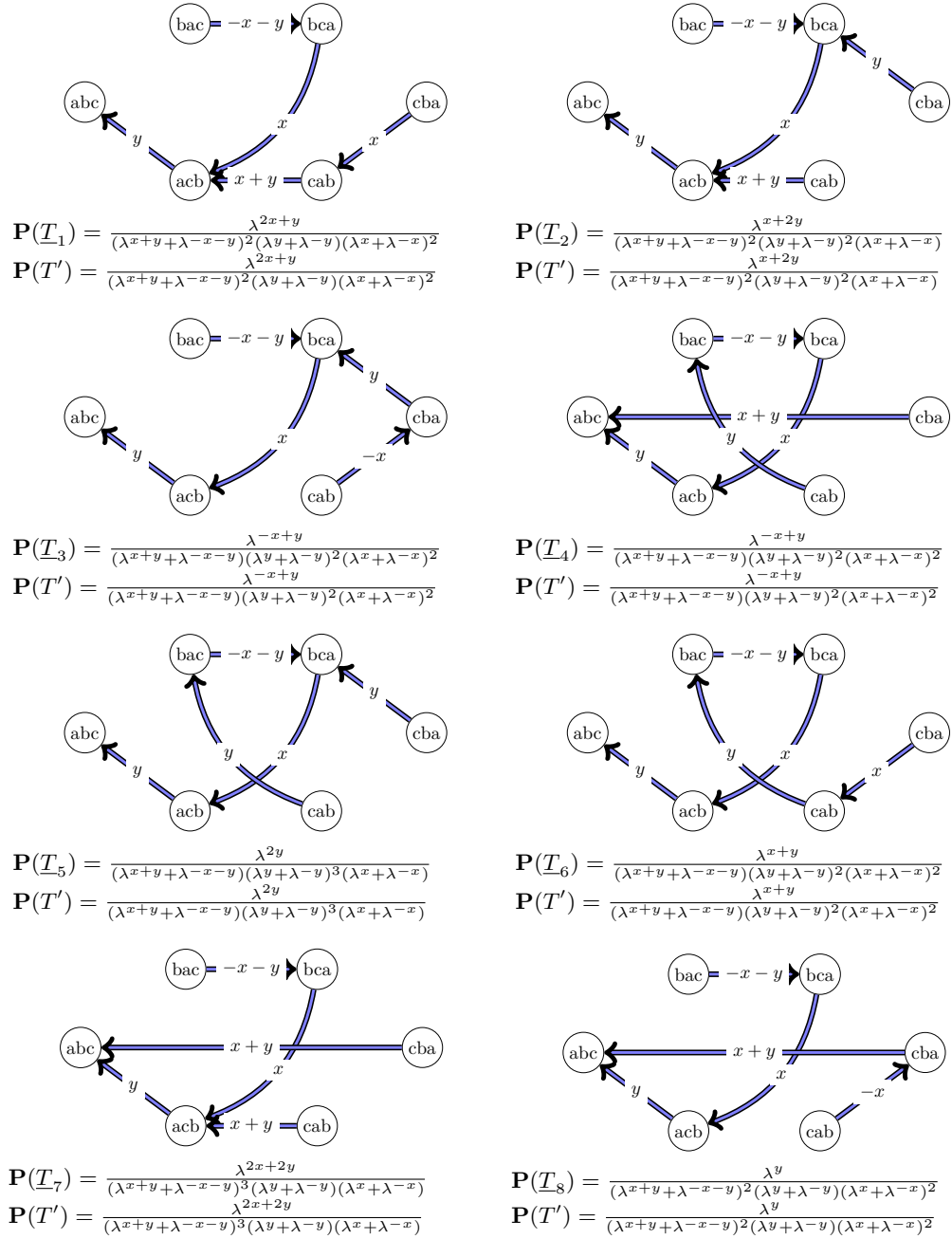

	\begin{figure}[h]
		\centering
		\begin{minipage}[b]{0.49\textwidth}
			\centering
			\resizebox{.8\linewidth}{!}{
				\begin{tikzpicture}
				\states
				
				\xbacbca \xbcacba \xcbaabc \xcabbac \xacbabc
				\end{tikzpicture}
			}  
			\label{fig:good9}
			\subcaption*{$\P(\underline T_9) = \frac{\lambda^{y}}{(\lambda^{x+y}+\lambda^{-x-y})^2(\lambda^{y}+\lambda^{-y})^3}$\\
				$\P(T') = \frac{\lambda^{3y}}{(\lambda^{x+y}+\lambda^{-x-y})^2(\lambda^{y}+\lambda^{-y})^3}$\\}
		\end{minipage}
		\begin{minipage}[b]{0.49\textwidth}
			\centering
			\resizebox{.8\linewidth}{!}{
				\begin{tikzpicture}
				\states
				
				\xbacbca \xbcacba \xcbaabc \xcabbac \xacbbca
				\end{tikzpicture}
			}   
			\label{fig:good10}
			\subcaption*{$\P(T) = \frac{\lambda^{-x}}{(\lambda^{x+y}+\lambda^{-x-y})^2(\lambda^{y}+\lambda^{-y})^2(\lambda^{x}+\lambda^{-x})}$\\
				$\P(T') = \frac{\lambda^{-x+2y}}{(\lambda^{x+y}+\lambda^{-x-y})^2(\lambda^{y}+\lambda^{-y})^2(\lambda^{x}+\lambda^{-x})}$\\}
		\end{minipage}
		
		\begin{minipage}[b]{0.49\textwidth}
			\centering
			\resizebox{.8\linewidth}{!}{
				\begin{tikzpicture}
				\states
				
				\xbacbca \xbcacba \xcbaabc \xcabbac \xacbcab
				\end{tikzpicture}
			}   
			\label{fig:good11}
			\subcaption*{$\P(T) = \frac{\lambda^{-x-y}}{(\lambda^{x+y}+\lambda^{-x-y})^3(\lambda^{y}+\lambda^{-y})^2}$\\
				$\P(T') = \frac{\lambda^{-x+y}}{(\lambda^{x+y}+\lambda^{-x-y})^3(\lambda^{y}+\lambda^{-y})^2}$\\}
		\end{minipage}
		\begin{minipage}[b]{0.49\textwidth}
			\centering
			\resizebox{.8\linewidth}{!}{
				\begin{tikzpicture}
				\states
				
				\xbacbca \xbcacba \xcbaabc \xacbabc \xcabcba
				\end{tikzpicture}
			}   
			\label{fig:good12}
			\subcaption*{$\P(T) = \frac{\lambda^{-x}}{(\lambda^{x+y}+\lambda^{-x-y})^2(\lambda^{y}+\lambda^{-y})^2(\lambda^{x}+\lambda^{-x})}$\\
				$\P(T') = \frac{\lambda^{-x+2y}}{(\lambda^{x+y}+\lambda^{-x-y})^2(\lambda^{y}+\lambda^{-y})^2(\lambda^{x}+\lambda^{-x})}$\\}
		\end{minipage}		
		
		\begin{minipage}[b]{0.49\textwidth}
			\centering
			\resizebox{.8\linewidth}{!}{
				\begin{tikzpicture}
				\states
				
				\xbacbca \xbcacba \xcbaabc \xacbbca \xcabcba
				\end{tikzpicture}
			}   
			\label{fig:good13}
			\subcaption*{$\P(T) = \frac{\lambda^{-2x-y}}{(\lambda^{x+y}+\lambda^{-x-y})^2(\lambda^{y}+\lambda^{-y})(\lambda^{x}+\lambda^{-x})^2}$\\
				$\P(T') = \frac{\lambda^{-2x+y}}{(\lambda^{x+y}+\lambda^{-x-y})^2(\lambda^{y}+\lambda^{-y})(\lambda^{x}+\lambda^{-x})^2}$\\}
		\end{minipage}
		\begin{minipage}[b]{0.49\textwidth}
			\centering
			\resizebox{.8\linewidth}{!}{
				\begin{tikzpicture}
				\states
				
				\xbacbca \xbcacba \xcbaabc \xacbcab \xcabcba
				\end{tikzpicture}
			}   
			\label{fig:good14}
			\subcaption*{$\P(T) = \frac{\lambda^{-2x-2y}}{(\lambda^{x+y}+\lambda^{-x-y})^3(\lambda^{y}+\lambda^{-y})(\lambda^{x}+\lambda^{-x})}$\\
				$\P(T') = \frac{\lambda^{-2x}}{(\lambda^{x+y}+\lambda^{-x-y})^3(\lambda^{y}+\lambda^{-y})(\lambda^{x}+\lambda^{-x})}$\\}
		\end{minipage}	
		
		\begin{minipage}[b]{0.49\textwidth}
			\centering
			\resizebox{.8\linewidth}{!}{
				\begin{tikzpicture}
				\states
				
				\xbacbca \xbcacba \xcbaabc \xacbabc \xcabacb
				\end{tikzpicture}
			}   
			\label{fig:good15}
			\subcaption*{$\P(T) = \frac{\lambda^{x+y}}{(\lambda^{x+y}+\lambda^{-x-y})^3(\lambda^{y}+\lambda^{-y})^2}$\\
				$\P(T') = \frac{\lambda^{x+3y}}{(\lambda^{x+y}+\lambda^{-x-y})^3(\lambda^{y}+\lambda^{-y})^2}$\\}
		\end{minipage}
		\begin{minipage}[b]{0.49\textwidth}
			\centering
			\resizebox{.8\linewidth}{!}{
				\begin{tikzpicture}
				\states
				
				\xbacbca \xbcacba \xcbaabc \xacbbca \xcabacb
				\end{tikzpicture}
			}   
			\label{fig:good16}
			\subcaption*{$\P(T) = \frac{\lambda^{0}}{(\lambda^{x+y}+\lambda^{-x-y})^3(\lambda^{y}+\lambda^{-y})(\lambda^{x}+\lambda^{-x})}$\\
				$\P(T') = \frac{\lambda^{2y}}{(\lambda^{x+y}+\lambda^{-x-y})^3(\lambda^{y}+\lambda^{-y})(\lambda^{x}+\lambda^{-x})}$\\}
		\end{minipage}		
		\caption{Good trees with path \ref{fig:bacabc:6}.}
		\label{fig:goodtreesF}
	\end{figure}
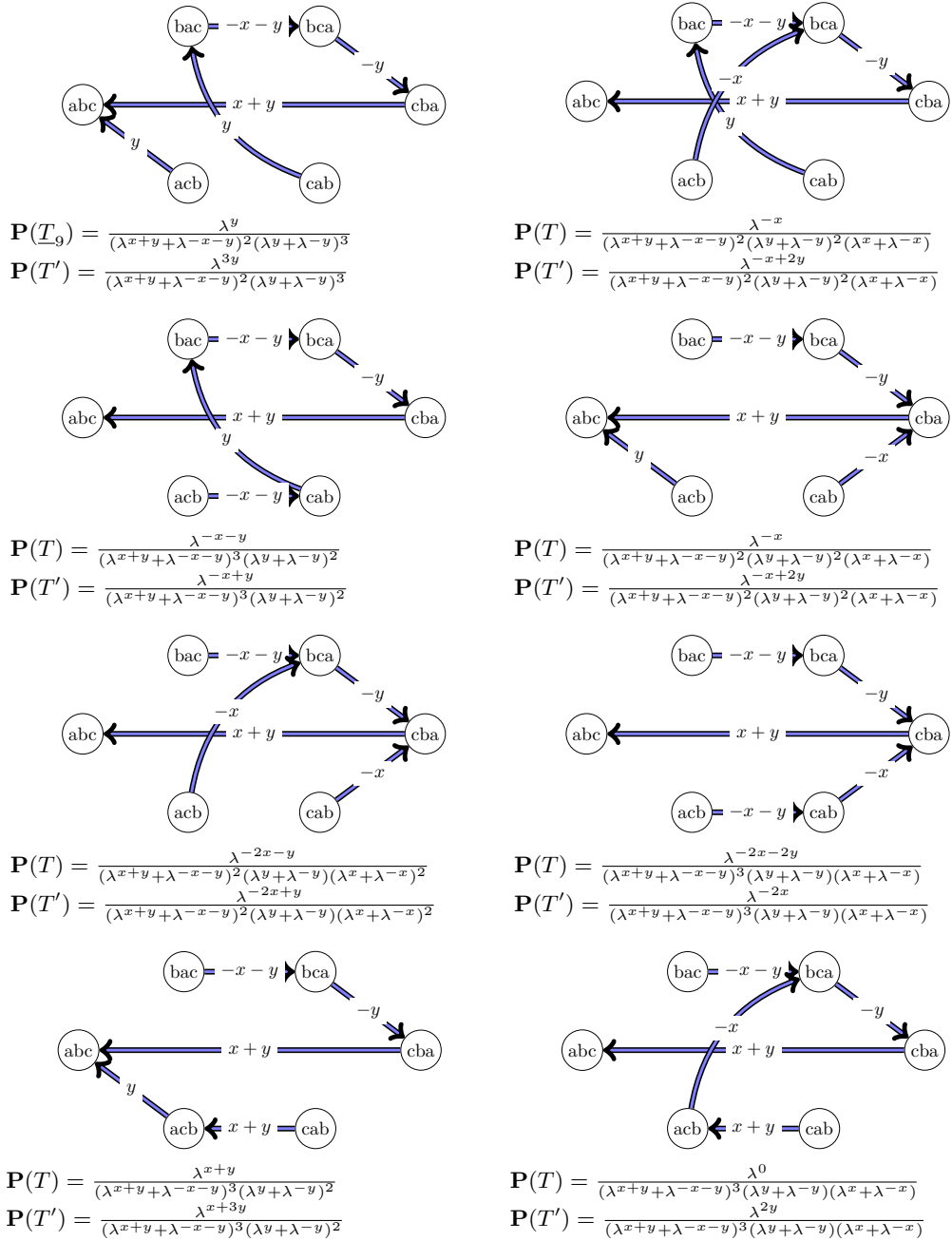

	\begin{figure}[h]
		\centering
		\begin{minipage}[b]{0.49\textwidth}
			\centering
			\resizebox{.8\linewidth}{!}{
				\begin{tikzpicture}
				\states
				
				\xbaccab \xcabcba \xcbaabc \xbcabac \xacbabc
				\end{tikzpicture}
			}  
			\label{fig:good17}
			\subcaption*{$\P(T) = \frac{\lambda^{x+2y}}{(\lambda^{x+y}+\lambda^{-x-y})^2(\lambda^{y}+\lambda^{-y})^2(\lambda^{x}+\lambda^{-x})}$\\
				$\P(T') = \frac{\lambda^{x+2y}}{(\lambda^{x+y}+\lambda^{-x-y})^2(\lambda^{y}+\lambda^{-y})^2(\lambda^{x}+\lambda^{-x})}$\\}
		\end{minipage}
		\begin{minipage}[b]{0.49\textwidth}
			\centering
			\resizebox{.8\linewidth}{!}{
				\begin{tikzpicture}
				\states
				
				\xbaccab \xcabcba \xcbaabc \xbcabac \xacbbca
				\end{tikzpicture}
			}   
			\label{fig:good18}
			\subcaption*{$\P(T) = \frac{\lambda^{y}}{(\lambda^{x+y}+\lambda^{-x-y})^2(\lambda^{y}+\lambda^{-y})(\lambda^{x}+\lambda^{-x})^2}$\\
				$\P(T') = \frac{\lambda^{y}}{(\lambda^{x+y}+\lambda^{-x-y})^2(\lambda^{y}+\lambda^{-y})(\lambda^{x}+\lambda^{-x})^2}$\\}
		\end{minipage}
		
		\begin{minipage}[b]{0.49\textwidth}
			\centering
			\resizebox{.8\linewidth}{!}{
				\begin{tikzpicture}
				\states
				
				\xbaccab \xcabcba \xcbaabc \xbcabac \xacbcab
				\end{tikzpicture}
			}   
			\label{fig:good19}
			\subcaption*{$\P(T) = \frac{\lambda^{0}}{(\lambda^{x+y}+\lambda^{-x-y})^3(\lambda^{y}+\lambda^{0})(\lambda^{x}+\lambda^{-x})}$\\
				$\P(T') = \frac{\lambda^{0}}{(\lambda^{x+y}+\lambda^{-x-y})^3(\lambda^{y}+\lambda^{-y})(\lambda^{x}+\lambda^{-x})}$\\}
		\end{minipage}
		\begin{minipage}[b]{0.49\textwidth}
			\centering
			\resizebox{.8\linewidth}{!}{
				\begin{tikzpicture}
				\states
				
				\xbaccab \xcabcba \xcbaabc \xbcaacb \xacbabc
				\end{tikzpicture}
			}   
			\label{fig:good20}
			\subcaption*{$\P(T) = \frac{\lambda^{x+y}}{(\lambda^{x+y}+\lambda^{-x-y})(\lambda^{y}+\lambda^{-y})^2(\lambda^{x}+\lambda^{-x})^2}$\\
				$\P(T') = \frac{\lambda^{x+y}}{(\lambda^{x+y}+\lambda^{-x-y})(\lambda^{y}+\lambda^{-y})^2(\lambda^{x}+\lambda^{-x})^2}$\\}
		\end{minipage}		
		
		\begin{minipage}[b]{0.49\textwidth}
			\centering
			\resizebox{.8\linewidth}{!}{
				\begin{tikzpicture}
				\states
				
				\xbaccab \xcabcba \xcbaabc \xbcaacb \xacbcab
				\end{tikzpicture}
			}   
			\label{fig:good21}
			\subcaption*{$\P(T) = \frac{\lambda^{-y}}{(\lambda^{x+y}+\lambda^{-x-y})^2(\lambda^{y}+\lambda^{-y})(\lambda^{x}+\lambda^{-x})^2}$\\
				$\P(T') = \frac{\lambda^{-y}}{(\lambda^{x+y}+\lambda^{-x-y})^2(\lambda^{y}+\lambda^{-y})(\lambda^{x}+\lambda^{-x})^2}$\\}
		\end{minipage}
		\begin{minipage}[b]{0.49\textwidth}
			\centering
			\resizebox{.8\linewidth}{!}{
				\begin{tikzpicture}
				\states
				
				\xbaccab \xcabcba \xcbaabc \xbcacba \xacbabc
				\end{tikzpicture}
			}   
			\label{fig:good22}
			\subcaption*{$\P(T) = \frac{\lambda^{0}}{(\lambda^{x+y}+\lambda^{-x-y})(\lambda^{y}+\lambda^{-y})^3(\lambda^{x}+\lambda^{-x})}$\\
				$\P(T') = \frac{\lambda^{0}}{(\lambda^{x+y}+\lambda^{-x-y})(\lambda^{y}+\lambda^{-y})^3(\lambda^{x}+\lambda^{-x})}$\\}
		\end{minipage}	
		
		\begin{minipage}[b]{0.49\textwidth}
			\centering
			\resizebox{.8\linewidth}{!}{
				\begin{tikzpicture}
				\states
				
				\xbaccab \xcabcba \xcbaabc \xbcacba \xacbbca
				\end{tikzpicture}
			}   
			\label{fig:good23}
			\subcaption*{$\P(T) = \frac{\lambda^{-x-y}}{(\lambda^{x+y}+\lambda^{-x-y})(\lambda^{y}+\lambda^{-y})^2(\lambda^{x}+\lambda^{-x})^2}$\\
				$\P(T') = \frac{\lambda^{-x-y}}{(\lambda^{x+y}+\lambda^{-x-y})(\lambda^{y}+\lambda^{-y})^2(\lambda^{x}+\lambda^{-x})^2}$\\}
		\end{minipage}
		\begin{minipage}[b]{0.49\textwidth}
			\centering
			\resizebox{.8\linewidth}{!}{
				\begin{tikzpicture}
				\states
				
				\xbaccab \xcabcba \xcbaabc \xbcacba \xacbcab
				\end{tikzpicture}
			}   
			\label{fig:good24}
			\subcaption*{$\P(T) = \frac{\lambda^{-x-2y}}{(\lambda^{x+y}+\lambda^{-x-y})^2(\lambda^{y}+\lambda^{-y})^2(\lambda^{x}+\lambda^{-x})}$\\
				$\P(T') = \frac{\lambda^{-x-2y}}{(\lambda^{x+y}+\lambda^{-x-y})^2(\lambda^{y}+\lambda^{-y})^2(\lambda^{x}+\lambda^{-x})}$\\}
		\end{minipage}		
		\caption{Good trees with path \ref{fig:bacabc:8}.}
		\label{fig:goodtreesH}
	\end{figure}
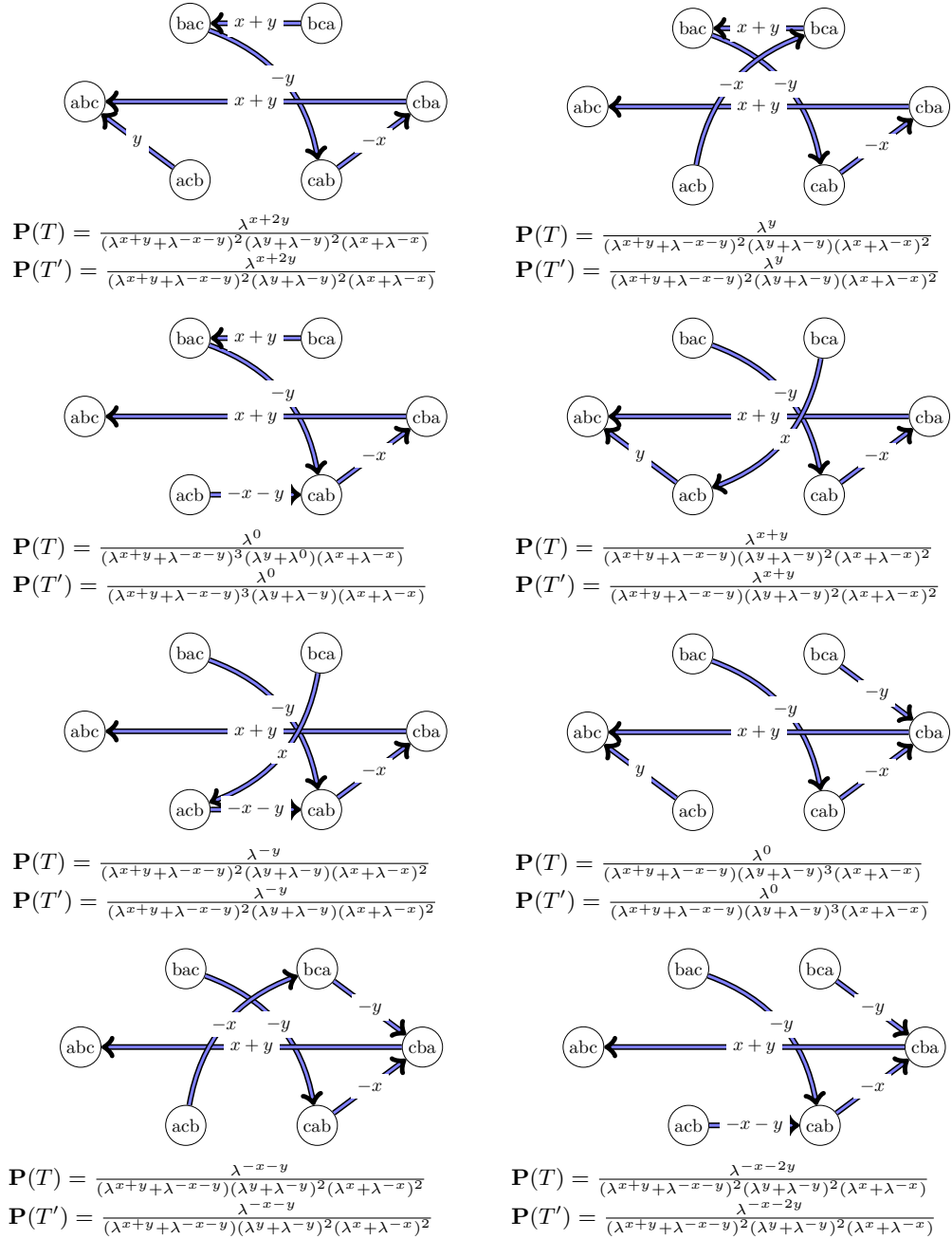

\begin{figure}[h]
	\begin{subfigure}[b]{0.32\textwidth}
		\centering
		\resizebox{\linewidth}{!}{
			\begin{tikzpicture}
			\states
			
			\cbabca
			\end{tikzpicture}
		}
		\caption{$\lambda^{c-b}$}
		\label{fig:bcacba:1}
	\end{subfigure}
	\begin{subfigure}[b]{0.32\textwidth}
		\centering
		\resizebox{\linewidth}{!}{
			\begin{tikzpicture}
			\states
			
			\cbacab \cabacb \acbabc \abcbac \bacbca
			\end{tikzpicture}
		}
		\caption{$\lambda^{c-b}$}   
		\label{fig:bcacba:2}
	\end{subfigure}
	\begin{subfigure}[b]{0.32\textwidth}
		\centering
		\resizebox{\linewidth}{!}{
			\begin{tikzpicture}
			\states
			
			\cbacab \cabbac \bacbca
			\end{tikzpicture}
		}
		\caption{$\lambda^{0}$}
		\label{fig:bcacba:3}
	\end{subfigure}
	
	\begin{subfigure}[b]{0.32\textwidth}
		\centering
		\resizebox{\linewidth}{!}{
			\begin{tikzpicture}
			\states
			
			\cbacab \cabbac \bacabc \abcacb \acbbca
			\end{tikzpicture}
		}
		\caption{$\lambda^{b-a}$}
		\label{fig:bcacba:4}
	\end{subfigure}
	\begin{subfigure}[b]{0.32\textwidth}
		\centering
		\resizebox{\linewidth}{!}{
			\begin{tikzpicture}
			\states
			
			\cbacab \cabacb \acbbca
			\end{tikzpicture}
		}
		\caption{$\lambda^{c-a}$}
		\label{fig:bcacba:5}
	\end{subfigure}
	\begin{subfigure}[b]{0.32\textwidth}
		\centering
		\resizebox{\linewidth}{!}{
			\begin{tikzpicture}
			\states
			
			\cbaabc \abcbac \bacbca
			\end{tikzpicture}
		}
		\caption{$\lambda^{a-b}$}
		\label{fig:bcacba:6}
	\end{subfigure}
	\begin{subfigure}[b]{0.32\textwidth}
		\centering
		\resizebox{\linewidth}{!}{
			\begin{tikzpicture}
			\states
			
			\cbaabc \abcbac \baccab \cabacb \acbbca
			\end{tikzpicture}
		}
		\caption{$\lambda^{c-b}$}
		\label{fig:bcacba:7}
	\end{subfigure}
	\begin{subfigure}[b]{0.32\textwidth}
		\centering
		\resizebox{\linewidth}{!}{
			\begin{tikzpicture}
			\states
			
			\cbaabc \abcacb \acbcab \cabbac \bacbca
			\end{tikzpicture}
		}
		\caption{$\lambda^{a-c}$}
		\label{fig:bcacba:8}
	\end{subfigure}
	\begin{subfigure}[b]{0.32\textwidth}
		\centering
		\resizebox{\linewidth}{!}{
			\begin{tikzpicture}
			\states
			
			\cbaabc \abcacb \acbbca
			\end{tikzpicture}
		}
		\caption{$\lambda^{0}$}
		\label{fig:bcacba:9}
	\end{subfigure}
	
	\caption{All paths from (cba) to (bca). Bad trees include path \subref{fig:bcacba:4} or \subref{fig:bcacba:5}.}
	\label{fig:bcacba}
\end{figure}


\begin{figure}[h]
	\begin{subfigure}[b]{0.32\textwidth}
		\centering
		\resizebox{\linewidth}{!}{
			\begin{tikzpicture}
			\states
			
			\acbabc
			\end{tikzpicture}
		}
		\caption{$\lambda^{c-b}$}
		\label{fig:acbabc:1}
	\end{subfigure}
	\begin{subfigure}[b]{0.32\textwidth}
		\centering
		\resizebox{\linewidth}{!}{
			\begin{tikzpicture}
			\states
			
			\acbcab \cabcba \cbabca \bcabac \bacabc
			\end{tikzpicture}
		}
		\caption{$\lambda^{c-b}$}   
		\label{fig:acbabc:2}
	\end{subfigure}
	\begin{subfigure}[b]{0.32\textwidth}
		\centering
		\resizebox{\linewidth}{!}{
			\begin{tikzpicture}
			\states
			
			\acbcab \cabbac \bacabc
			\end{tikzpicture}
		}
		\caption{$\lambda^{0}$}
		\label{fig:acbabc:3}
	\end{subfigure}
	
	\begin{subfigure}[b]{0.32\textwidth}
		\centering
		\resizebox{\linewidth}{!}{
			\begin{tikzpicture}
			\states
			
			\acbbca \bcacba \cbacab \cabbac \bacabc 
			\end{tikzpicture}
		}
		\caption{$\lambda^{b-a}$}
		\label{fig:acbabc:4}
	\end{subfigure}
	\begin{subfigure}[b]{0.32\textwidth}
		\centering
		\resizebox{\linewidth}{!}{
			\begin{tikzpicture}
			\states
			
			\acbbca \bcabac \bacabc
			\end{tikzpicture}
		}
		\caption{$\lambda^{c-a}$}
		\label{fig:acbabc:5}
	\end{subfigure}
	\begin{subfigure}[b]{0.32\textwidth}
		\centering
		\resizebox{\linewidth}{!}{
			\begin{tikzpicture}
			\states
			
			\acbcab \cabcba \cbaabc
			\end{tikzpicture}
		}
		\caption{$\lambda^{a-b}$}
		\label{fig:acbabc:6}
	\end{subfigure}
	\begin{subfigure}[b]{0.32\textwidth}
		\centering
		\resizebox{\linewidth}{!}{
			\begin{tikzpicture}
			\states
			
			\acbbca \bcabac \baccab \cabcba \cbaabc
			\end{tikzpicture}
		}
		\caption{$\lambda^{c-b}$}
		\label{fig:acbabc:7}
	\end{subfigure}
	\begin{subfigure}[b]{0.32\textwidth}
		\centering
		\resizebox{\linewidth}{!}{
			\begin{tikzpicture}
			\states
			
			\acbcab \cabbac \bacbca \bcacba \cbaabc
			\end{tikzpicture}
		}
		\caption{$\lambda^{a-c}$}
		\label{fig:acbabc:8}
	\end{subfigure}
	\begin{subfigure}[b]{0.32\textwidth}
		\centering
		\resizebox{\linewidth}{!}{
			\begin{tikzpicture}
			\states
			
			\cbaabc \bcacba \acbbca
			\end{tikzpicture}
		}
		\caption{$\lambda^{0}$}
		\label{fig:acbabc:9}
	\end{subfigure}
	
	\caption{All paths from (acb) to (abc). Bad trees include path \subref{fig:acbabc:4} or \subref{fig:acbabc:5}.} 
	\label{fig:acbabc}
\end{figure}

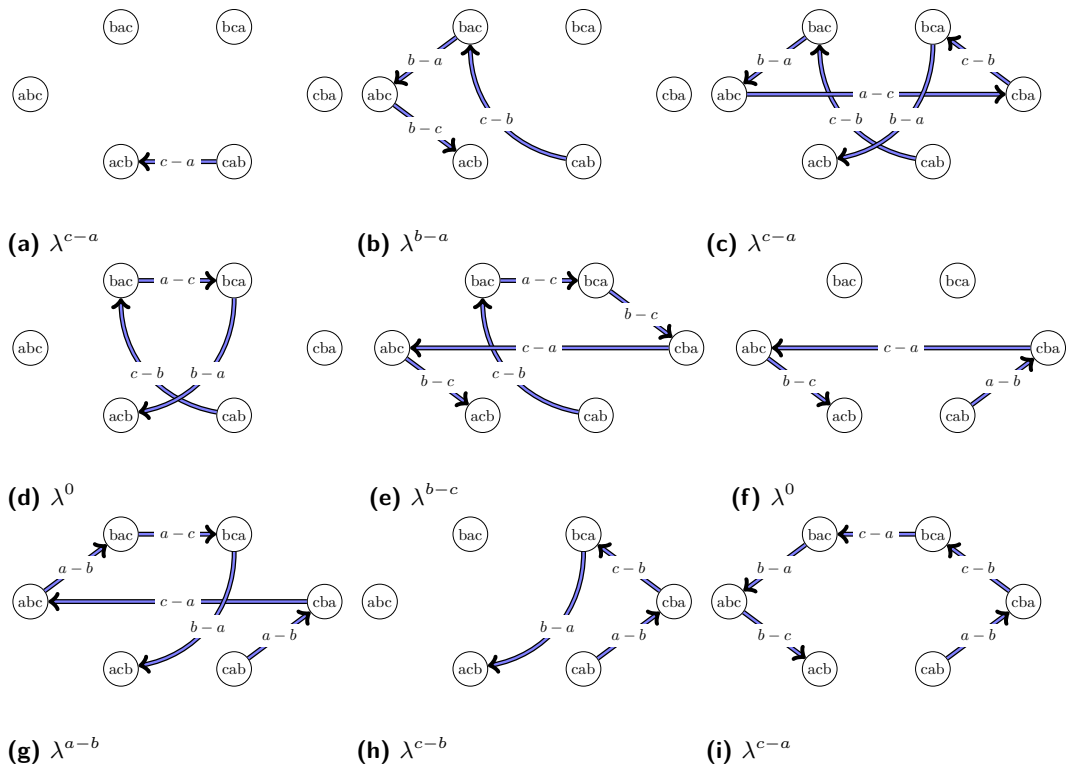
\begin{figure}[h]
	\begin{subfigure}[b]{0.32\textwidth}
		\centering
		\resizebox{\linewidth}{!}{
			\begin{tikzpicture}
			\states
			
			\cabacb
			\end{tikzpicture}
		}
		\caption{$\lambda^{c-a}$}
		\label{fig:cabacb:1}
	\end{subfigure}
	\begin{subfigure}[b]{0.32\textwidth}
		\centering
		\resizebox{\linewidth}{!}{
			\begin{tikzpicture}
			\states
			
			\cabbac \bacabc \abcacb
			\end{tikzpicture}
		}
		\caption{$\lambda^{b-a}$}   
		\label{fig:cabacb:2}
	\end{subfigure}
	\begin{subfigure}[b]{0.32\textwidth}
		\centering
		\resizebox{\linewidth}{!}{
			\begin{tikzpicture}
			\states
			
			\cabbac \bacabc \abccba \cbabca \bcaacb
			\end{tikzpicture}
		}
		\caption{$\lambda^{c-a}$}
		\label{fig:cabacb:3}
	\end{subfigure}
	
	\begin{subfigure}[b]{0.32\textwidth}
		\centering
		\resizebox{\linewidth}{!}{
			\begin{tikzpicture}
			\states
			
			\cabbac \bacbca \bcaacb
			\end{tikzpicture}
		}
		\caption{$\lambda^{0}$}
		\label{fig:cabacb:4}
	\end{subfigure}
	\begin{subfigure}[b]{0.32\textwidth}
		\centering
		\resizebox{\linewidth}{!}{
			\begin{tikzpicture}
			\states
			
			\cabbac \bacbca \bcacba \cbaabc \abcacb
			\end{tikzpicture}
		}
		\caption{$\lambda^{b-c}$}
		\label{fig:cabacb:5}
	\end{subfigure}
	\begin{subfigure}[b]{0.32\textwidth}
		\centering
		\resizebox{\linewidth}{!}{
			\begin{tikzpicture}
			\states
			
			\cabcba \cbaabc \abcacb
			\end{tikzpicture}
		}
		\caption{$\lambda^{0}$}
		\label{fig:cabacb:6}
	\end{subfigure}
	\begin{subfigure}[b]{0.32\textwidth}
		\centering
		\resizebox{\linewidth}{!}{
			\begin{tikzpicture}
			\states
			
			\cabcba \cbaabc \abcbac \bacbca \bcaacb
			\end{tikzpicture}
		}
		\caption{$\lambda^{a-b}$}
		\label{fig:cabacb:7}
	\end{subfigure}
	\begin{subfigure}[b]{0.32\textwidth}
		\centering
		\resizebox{\linewidth}{!}{
			\begin{tikzpicture}
			\states
			
			\cabcba \cbabca \bcaacb
			\end{tikzpicture}
		}
		\caption{$\lambda^{c-b}$}
		\label{fig:cabacb:8}
	\end{subfigure}
	\begin{subfigure}[b]{0.32\textwidth}
		\centering
		\resizebox{\linewidth}{!}{
			\begin{tikzpicture}
			\states
			
			\cabcba \cbabca \bcabac \bacabc \abcacb
			\end{tikzpicture}
		}
		\caption{$\lambda^{c-a}$}
		\label{fig:cabacb:9}
	\end{subfigure}
	
	\caption{All paths from (cab) to (acb). There are no bad trees for these two states.} 
	\label{fig:cabacb}
\end{figure}

\begin{figure}[h]
	\begin{subfigure}[b]{0.32\textwidth}
		\centering
		\resizebox{\linewidth}{!}{
			\begin{tikzpicture}
			\states
			
			\cbacab
			\end{tikzpicture}
		}
		\caption{$\lambda^{b-a}$}
		\label{fig:cbacab:1}
	\end{subfigure}
	\begin{subfigure}[b]{0.32\textwidth}
		\centering
		\resizebox{\linewidth}{!}{
			\begin{tikzpicture}
			\states
			
			\cbabca \bcaacb \acbcab
			\end{tikzpicture}
		}
		\caption{$\lambda^{0}$}   
		\label{fig:cbacab:2}
	\end{subfigure}
	\begin{subfigure}[b]{0.32\textwidth}
		\centering
		\resizebox{\linewidth}{!}{
			\begin{tikzpicture}
			\states
			
			\cbabca \bcaacb \acbabc \abcbac \baccab
			\end{tikzpicture}
		}
		\caption{$\lambda^{c-b}$}
		\label{fig:cbacab:3}
	\end{subfigure}
	
	\begin{subfigure}[b]{0.32\textwidth}
		\centering
		\resizebox{\linewidth}{!}{
			\begin{tikzpicture}
			\states
			
			\cbabca \bcabac \baccab
			\end{tikzpicture}
		}
		\caption{$\lambda^{c-a}$}
		\label{fig:cbacab:4}
	\end{subfigure}
	\begin{subfigure}[b]{0.32\textwidth}
		\centering
		\resizebox{\linewidth}{!}{
			\begin{tikzpicture}
			\states
			
			\cbabca \bcabac \bacabc \abcacb \acbcab
			\end{tikzpicture}
		}
		\caption{$\lambda^{b-a}$}
		\label{fig:cbacab:5}
	\end{subfigure}
	\begin{subfigure}[b]{0.32\textwidth}
		\centering
		\resizebox{\linewidth}{!}{
			\begin{tikzpicture}
			\states
			
			\cbaabc \abcacb \acbcab
			\end{tikzpicture}
		}
		\caption{$\lambda^{b-c}$}
		\label{fig:cbacab:6}
	\end{subfigure}
	\begin{subfigure}[b]{0.32\textwidth}
		\centering
		\resizebox{\linewidth}{!}{
			\begin{tikzpicture}
			\states
			
			\cbaabc \abcacb \acbbca \bcabac \baccab
			\end{tikzpicture}
		}
		\caption{$\lambda^{b-a}$}
		\label{fig:cbacab:7}
	\end{subfigure}
	\begin{subfigure}[b]{0.32\textwidth}
		\centering
		\resizebox{\linewidth}{!}{
			\begin{tikzpicture}
			\states
			
			\cbaabc \abcbac \baccab
			\end{tikzpicture}
		}
		\caption{$\lambda^{0}$}
		\label{fig:cbacab:8}
	\end{subfigure}
	\begin{subfigure}[b]{0.32\textwidth}
		\centering
		\resizebox{\linewidth}{!}{
			\begin{tikzpicture}
			\states
			
			\cbaabc \abcbac \bacbca \bcaacb \acbcab
			\end{tikzpicture}
		}
		\caption{$\lambda^{a-c}$}
		\label{fig:cbacab:9}
	\end{subfigure}
	
	\caption{All paths from (cba) to (cab). Bad trees include path \subref{fig:cbacab:3} or \subref{fig:cbacab:4}.} 
	\label{fig:cbacab}
\end{figure}

\end{document}